\documentclass[trackchanges,twocolumn,tighten,mathlines]{aastex7}
\usepackage{amsmath}

\begin{document}

\title{Morphologies of SAGAbg low-mass galaxies in Legacy Survey multi-band imaging:\\ dependence on stellar masses, star-formation rates and low-redshift evolution}

\author[orcid=0000-0001-6442-5786,sname='Joy Bhattacharyya']{Joy Bhattacharyya}
\affiliation{Department of Physics \& Astronomy, Amherst College, 6 East Drive, Amherst, MA 01002, USA}
\affiliation{Department of Astronomy, The Ohio State University, Columbus, OH 43210, USA}
\affiliation{Center for Cosmology and Astro-Particle Physics, The Ohio State University, Columbus, OH 43210, USA}
\email[show]{jbhattacharyya@amherst.edu} 
\author[orcid=0000-0003-4394-7491]{Guinevere Herron}
\affiliation{Department of Physics, Dartmouth College, Hanover, NH 03755, USA}
\email{guinevere.herron.GR@dartmouth.edu}
\author[orcid=0000-0002-8320-2198]{Yasmeen Asali}
\affiliation{Department of Astronomy, Yale University, New Haven, CT 06520, USA}
\email{xx}
\author[orcid=0000-0002-9816-9300]{Abby Mintz}
\affiliation{Department of Astrophysical Sciences, Princeton University, 4 Ivy Lane, Princeton, NJ 08544, USA}
\email{xx}
\author[orcid=0000-0002-4739-046X]{Mithi A. C. de los Reyes}
\affiliation{Department of Physics \& Astronomy, Amherst College, 6 East Drive, Amherst, MA 01002, USA}
\email{xx}
\author[orcid=0000-0002-1200-0820]{Yao-Yuan Mao}
\affiliation{Department of Physics and Astronomy, University of Utah, Salt Lake City, UT 84112, USA}
\email{xx}
\author[orcid=0000-0002-8040-6785]{Annika H.G. Peter}
\affiliation{Department of Physics, The Ohio State University, Columbus, OH 43210, USA}
\affiliation{Center for Cosmology and Astro-Particle Physics, The Ohio State University, Columbus, OH 43210, USA}
\affiliation{Department of Astronomy, The Ohio State University, Columbus, OH 43210, USA}
\email{peter.33@osu.edu}

\shorttitle{Non-parametric morphologies of SAGAbg galaxies}
\shortauthors{Bhattacharyya et al.}

\begin{abstract}

The optical morphologies of low-mass galaxies can be used to directly trace their assembly and constrain models of galaxy evolution. We select a sample of 6211 low-mass ($7\lesssim {\rm log}(M_{\ast}/M_{\odot})\lesssim 10$) star-forming galaxies from the SAGAbg catalog that has high-completeness at low-redshifts ($z<0.1$). We obtain their galaxy maps in the $griz$ - bands of the Legacy Surveys and apply STATMORPH to calculate non-parametric morphological measures including the Gini index, $M_{20}$ measure, and CAS parameters. We study how resolution and signal-to-noise affect the morphology measures and find that the bulge strength measurements are the most reliable. The sequence in Gini$-M_{20}$ space is directly linked to the star-forming sequence of galaxies and is dominated by Sb/Sc/Ir morphologies. The $g$-band light distributions are the least concentrated among all the bands. The systematic trends of $M_{20}$ with respect to stellar mass and GALEX NUV-derived specific star formation rate (sSFR) strongly indicate that the galaxies with flatter light profiles are less massive with higher sSFR and vice-versa. We statistically infer that star-forming low-mass galaxies predominantly have disk morphologies with bulges becoming more prominent in quenched systems at higher masses (${\rm log}(M_{\ast}/M_{\odot})\gtrsim 9$).


\end{abstract}




\submitjournal{ApJ}

\section{Introduction} \label{sec:intro}

The term ``morphology" of a galaxy expresses its visual appearance that results from the structure of the constituent stars and gas. These structures, in different wavelengths, trace the physical properties of galaxies, and the evolution thereof. The mass, star-formation rate, age, metallicity and gas content are some of the properties connected to the thermal and dynamical histories of galaxies that contribute to their morphologies \citep{2025arXiv250209610M}. Taxonomic studies of galaxies based on their optical morphologies have been powerful, e.g., the subjective classifications based on the Hubble and de Vaucouleurs sequences \citep{1926ApJ....64..321H,1959HDP....53..275D} but with the large galaxy catalogs from modern wide-area surveys it has become necessary to circumvent human intervention in classification \citep{2008MNRAS.389.1179L}. Here, quantitative measures of a galaxy's morphology have emerged as a powerful probe of the shape, size, and concentration of the light profile. For example, modeling the light distribution as a S{\'e}rsic model \citep{1963BAAA....6...41S} has been applied for a wide range of galaxies. A range of non-parametric indices --- concentration, asymmetry, smoothness (CAS) \citep{2003ApJS..147....1C}, Gini-$M_{20}$ \citep{2004AJ....128..163L} --- have been used to measure higher-order moments, nonlinearities of the light-profile as well. Such methods have been further complemented with deep-learning models \citep[e.g.][]{2020A&C....3000334B,2020MNRAS.491.1554W,2022MNRAS.513.1581W,2023MNRAS.518.2794C}.



The Hubble and de Vaucouleurs schemes have established a paradigm in which these morphological sequences are analogous to the star-forming sequence with ``late-type" star-forming disky spiral galaxies evolving into the ``early-type" quenched elliptical galaxies \citep{1998ARA&A..36..189K}. The parametric and non-parametric approaches have further highlighted this connection \citep[e.g.][]{2003ApJS..147....1C,2006MNRAS.368..414D,Wuyts2011,2015Snyder} for massive galaxies. As a result the rich phenomenology of baryonic and dark matter physics that modify star formation and thereby the morphologies of the galaxies have been identified \citep{2009ApJS..182..216K}. Alongside star formation driven feedback, external processes including AGN feedback and major mergers, play a significant role in this regard \citep{2008ApJ...672..177L,2016MNRAS.463.3948D}. While the connection between the morphologies and evolution of massive galaxies have been comprehensively studied, our understanding of their low-mass contemporaries is only emerging \citep[e.g.][]{2020ApJ...900..163K,2020AJ....159..103K}.

Low-mass galaxies, i.e., with stellar masses of $7\lesssim {\rm log}(M_{\ast}/M_{\odot})\lesssim 10$, directly contribute to the mass assembly of massive galaxies like the MW \citep{2020ApJ...893...48N,2025ApJ...994...94T}. More so, these are present in abundance throughout the universe and also help us understand the evolution of high-redshift galaxies \citep{2021A&A...646A.138I,2022A&A...665L...4S,2025ApJ...987...90S}, that can be studied up-close. They are predominantly found to be star-forming in the field \citep[e.g.][]{Geha2012sdss,2014MNRAS.442.1396W,2015ApJ...804..136W}, with any deviations strongly depending on the galaxies' environments \citep{Bhattacharyya2025tng50}. The morphologies of these bright and classical dwarf galaxies can provide a more resolved understanding of their assembly history. The star-forming dwarfs are spheroidal, irregular or dispersion supported disks in contrast to their massive disk counterparts \citep{2006AJ....131..296S}. A consistent theory of galaxy evolution that can explain these phenomena across mass, environment and cosmic time is necessary.

Studies of low-mass galaxies in low-density environments highlight the significance of secular processes \citep{2023ApJ...951...52D}, as opposed to the environmental processes, i.e., mergers and tidal effects, that dominate in massive galaxies. Star formation drives bursty feedback inside the shallow potential wells of the low-mass halos causing the disks to puff up \citep{2010MNRAS.406L..65S,2016ApJ...820..131E}. More so, pseudo-bulges can develop in star-forming disks entirely through secular processes \citep{2004ARA&A..42..603K}. H$\alpha$ morphologies of low-mass galaxies further show that rapidly star-forming galaxies have compact morphologies with inhomogeneities due to clumpy star-forming regions \citep{2024ApJ...974..273M}. The non-parametric measures of morphologies in this regime have also been found to scale with respect to the galaxy's primary properties including mass \citep{Martin2025cosmic}.


Previous studies have often focused on small, incomplete samples and used single-band photometry to study the morphologies of dwarf galaxies. Therefore at this juncture we need to comprehensively study the morphologies of an extensive low-mass galaxy population in multi-band photometry. Extending the broad scope of this work we want to measure the morphologies and perform further analysis by taking into account auxiliary measurements that include stellar mass, star-formation rate and redshift evolution. For such a complex dataset, non-parametric morphologies, as measured by the STATMORPH code \citep{2019Rodriguez-Gomez} will be necessary as we statistically trace the relationship between galaxy properties and their morphologies. 

In our novel work, we select the sample of low-mass galaxies from the SAGAbg catalog at redshifts $z<0.1$ and with $r$-band magnitudes $18<r<20.75$. We detail our methodology in Sec. \ref{sec:methods} that involves the processing of Legacy Survey multi-band imaging, source detection and segmentation procedures, and application of STATMORPH. Among our results in Sec. \ref{sec:results}, we discuss the effect of systematics, analyze the dependence on stellar mass and sSFR, study the $Gini-M_{20}$ distributions and perform dimensionality reduction. We place our results in context with the established literature in morphology and galaxy evolution as well as cite the limitations arising due to systematics in Sec. \ref{sec:discussion}. In this work we use the \citet{2020A&A...641A...6P} cosmology, with $H_0=70$ Mpc km$^{-1}$ s and $\Omega_0=0.27$.

\section{Methods} \label{sec:methods}

\subsection{SAGAbg Sample} \label{sec:SAGAbg}

We use the SAGA Background Sample (SAGAbg) that is a by-product of the Satellites Around Galactic Analogs
(SAGA) Survey to collect spectroscopic redshifts of dwarf galaxies \citep{Geha2017saga1,Mao2024saga3}. More than 40000 spectra from archival SDSS/GAMA observations and obtained with AAT/2dF or MMT/Hectospec were taken within the targeting-complete footprint comprising projected distances $r<300$ kpc of MW-analog hosts at $z \sim 0.01$. The photometric selection employed in SAGA, down to a limiting magnitude
of $m_r \sim 21$, was chosen to target low-mass, low-redshift galaxies. This selection also favors the blue or star forming galaxies as opposed to their red or quenched counterparts. The SAGAbg sample is defined as those among these spectra that are not satellites of any of the SAGA hosts in \citet{Mao2024saga3}. Therefore this sample is predominantly made of dwarf galaxies that are not satellites of the SAGA host galaxies \citep{KadoFong2024sagabg1}.

The stellar mass for galaxies in the catalog was calculated as per the prescription in \citet{Mao2021saga2}, with uncertainties of $\sim 0.2$ dex \citep{Mao2024saga3},

\begin{equation} \label{eq:stellar_mass}
{\rm log}\biggl(\frac{\mathcal{M}_{\ast}}{M_{\odot}} \biggr)
= 1.254 + 1.098(g-r)_{0} - 0.4M_{r,0}.    
\end{equation}

\begin{figure}
\centering
\includegraphics[scale=0.3]{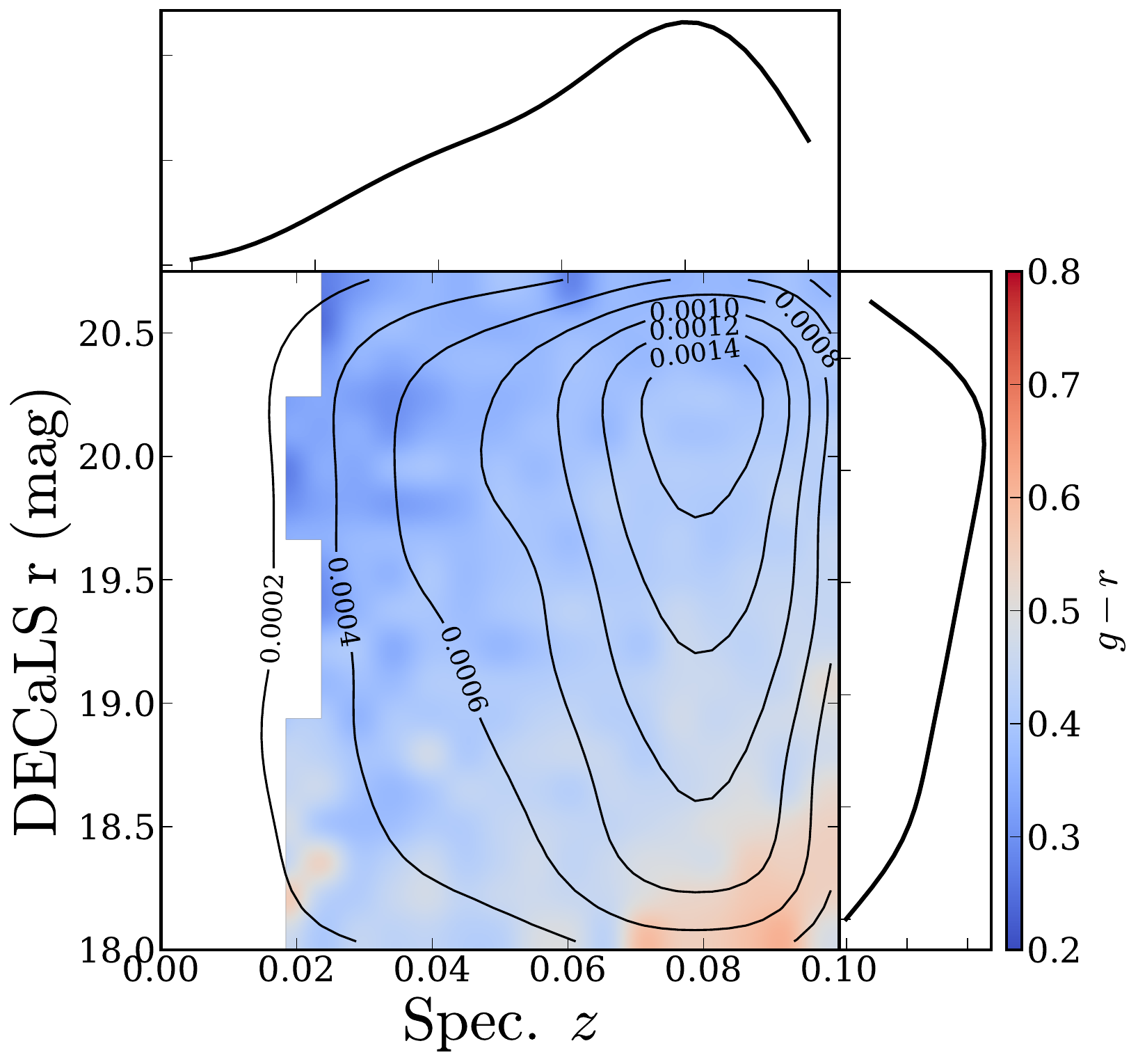}
\caption{The 2D distribution of SAGAbg-morph sample in the spectroscopic redshift–apparent magnitude plane. The main panel shows a Kernel Density Estimate of the galaxy distribution as a function of spectroscopic redshift ($z_{\rm spec}$) and Legacy Survey $r$-band magnitude, with contour levels indicating the density. The color scale, from blue to red, represents the median $g - r$ color for the sample binned in the grid. The top panel displays the marginalized redshift distribution, while the right panel shows the marginalized magnitude distribution. The distribution exhibits a broad peak around $z_{\rm spec} \approx 0.08$ and $r \approx 20.2$ mag \label{fig:selection1}}
\end{figure}

We select our sample by imposing the limits of $z<0.1$ and $18<m_r<20.75$ on the spectroscopic redshift and $r$-band given in the SAGAbg catalog respectively. While the magnitude limits straddle the typical selection of the SAGA sample, the redshift limit has been placed after considering both the issue of resolving sources at higher redshift as well as the inability to study redshift evolution of morphology/star-formation in a sample that is very localized \citep{KadoFong2024sagabg2}. This results in a sample of 8036 galaxies. In Fig. \ref{fig:selection1} we plot the selected sample in the space of $r$-band apparent magnitude and spectroscopic redshift. The median $g-r$ color of the galaxies in bins of $m_{r0}$ and $z_{\rm spec}$ are mapped out with the solid black contours showing the Kernel Density Estimate (KDE) of the galaxy distribution. The distribution exhibits a peak around $z_{\rm spec} \approx 0.08$ and $r \approx 20.2$ mag. 

\subsection{Legacy Survey Selection}

We use the selection of 8036 sources from the SAGAbg sample and run a query on NOIR Astro Data Lab \footnote{https://datalab.noirlab.edu} to match these to sources in the DESI Legacy Survey (LS) \citep{Dey2020LS} catalogs. The LS source catalogs are constructed using TRACTOR \citep{2016ascl.soft04008L} that performs source-extraction on the pixel-level data. We retrieve the TRACTOR based photometry and exclude sources that have bright stars/galaxies in their vicinity.  The majority of the SAGAbg sample is located in the southern sky with photometry from the DECam Legacy Survey (DECaLS) with the rest that lie in the northern sky that is covered by the Beijing-Arizona Sky Survey (BASS) and the Mayall z-band Legacy Survey (MzLS). The northern sky is defined as DEC$> 32.375^{\circ}$. While DECaLS comprises $griz$ band photometry, BASS/MzLS only has $grz$ bands. The center wavelengths /widths of the four broadband filters are- 4720/1520 $\mathring{A}$ for the $g$-band, 6415/1480 $\mathring{A}$ for the $r$-band, 7835/1470 $\mathring{A}$ for the $i$-band and 9260/1520 $\mathring{A}$ for the $z$-band.

We cross-match our selection from SAGAbg with the sources in the TRACTOR catalog for LS DR10. We identify the brightest source with ${\tt TYPE!=PSF}$, non-zero fluxes in the $gr$ bands and within 3$^{''}$ as the matching galaxy. We use the values of the given PSF FWHM sizes, ${\tt DCHISQ}$ parameters as well as calculate the magnitudes from the fluxes with reddening in each of the four bands. We also search for bright objects within 18$^{''}$ or the approximate size of the cutouts with $gr$ band magnitudes $<18$ mag and reject any source where we find bright neighbors as these image cutouts will have contamination. This leaves us with 6337 sources of which 4840 are in the DECaLS catalog and 1497 are in the BASS/MzLS catalog.

For each of these sources in the new catalog we download $128\times128$ pixel cutouts from the Legacy Survey server at the native pixel scale of 0.262 pixel arcsec$^{-1}$. For each of these galaxy maps we obtain the inverse variance maps as well. We perform image processing, segmentation and deblending that we outline in the following Sec. \ref{sec:image_proc} and \ref{sec:segment}. There are certain sources for which the pipeline fails to reliably perform these steps and we are left with 6211 galaxies that make up the SAGAbg-morph sample. This is the primary sample in our work and we use this to measure the morphologies using STATMORPH.

\begin{table}
    \centering
    \caption{Sizes of the various samples and sub-samples used in this work.}
    \begin{tabular}{l|cccc}
    \hline
    Catalog & & Size & & \\
    \hline
        SAGAbg in-band &  & 8036 &  & \\
        --LS match &  & 6337 &  & \\
        ---DECaLS DR10 &  & 4840  &  & \\
        ---BASS/MZLS DR9 &  & 1497 &  & \\
        ------SAGAbg-morph &  & 6211 &  & \\
        \hline
        -GALEX obs. &  & 6337 &  & \\
        --- S/N$>4$ &  & 3532 &  & \\
        --- S/N$<4$ &  & 2805 &  & \\
         \hline
    \end{tabular}
    
    \label{tab:placeholder}
\end{table}

\subsection{GALEX Image Processing \& SFR Measurement} \label{sec:galex}

\begin{figure*}
\centering
\includegraphics[scale=0.3]{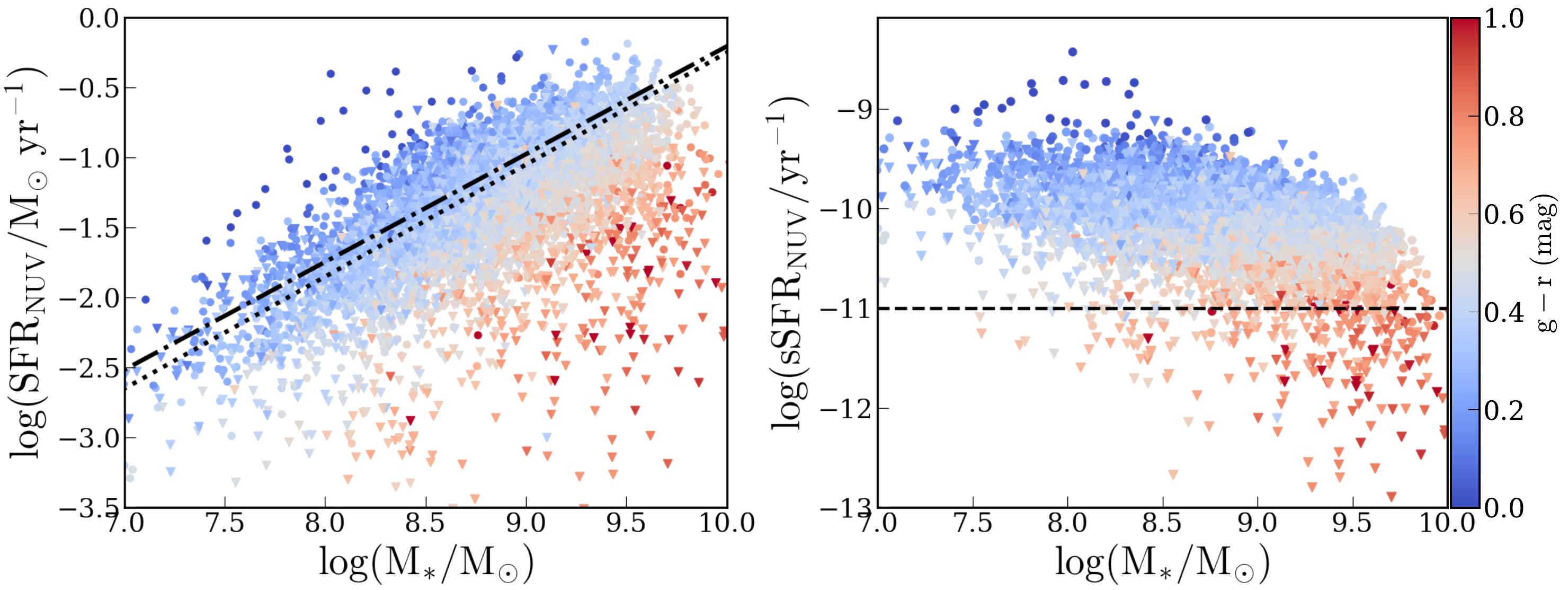}
\caption{Star-forming sequence (SFS) for the galaxies in the SAGAbg-morph sample. (\textit{Left}:) Star formation rate derived from GALEX NUV photometry versus stellar mass. The points represent measurements with S/N$>4$ whereas the upside-down triangles show the upper limit of the measurements with S/N$<4$. The markers are color-coded by the $g-r$ color from SAGAbs catalog, with blue galaxies tracing the star-forming sequence and red galaxies representing quiescent systems below the SFS. \textit{Dotted} and \textit{dash-dotted} lines shows the best-fits in \citet{KadoFong2024sagabg2} in the different redshift ranges: $0.0 < z_{\rm spec} < 0.035$ and $0.07 < z_{\rm spec} < 0.105$ respectively, illustrating the mild evolution of the SFS with redshift. (\textit{Right}:) Specific star formation rate (sSFR) as a function of stellar mass. The horizontal dot-dashed line marks ${\rm log}({\rm sSFR/yr}^{-1}) = -11.0$, a commonly used threshold separating star-forming from quiescent galaxies. Here we notice that there is an appreciable fraction of quenched galaxies at ${\rm log}(M_{\ast}/M_{\odot})\gtrsim 9$. \label{fig:sfms}}
\end{figure*}

An important galaxy property in the context of this work is the star-formation rate (SFR). To measure this we require complementary data from the GALEX UV survey \citep{2007ApJS..173..682M}. Near-UV (NUV) fluxes in GALEX trace the star-formation histories on timescales of the order of 100 Myr \citep{2009ApJ...706..599L,2019ApJ...873...74B}. We follow the prescription of \citet{Geha2024saga4} to measure the star-formation rates (SFR) of the sample using NUV flux measurements from GALEX. 

We download the $5' \times 5'$ cutouts in the GALEX NUV band for each of the 6337 SAGAbg sources matched to the LS database and mask out sources in the TRACTOR catalog with optical $m_{r0}<21$ mag. We estimate the background signal and RMS variance from the masked map. After background subtraction, we perform aperture photometry on the central source using an elliptical aperture having parameters determined from the SAGAbg catalog. We use a circular aperture of the size of the GALEX PSF FWHM$=6''$ if the optical radius is smaller than it. A sample of 3532 galaxies satisfy the threshold of S/N$>4$ as adopted in \citet{Geha2024saga4} while for 2805 galaxies with S/N$<4$ we assume that the flux measurements represent upper limits. We calculate the NUV luminosity from the the fluxes using the internal extinction correction made using the assumption of a \citet{2000ApJ...533..682C} reddening curve and $R_V = 3.67$ determined in nearby star-bursting galaxies \citep{2016ApJ...818...13B}. We proceed to estimate the SFR with the prescription of \citet{2006ApJS..164...38I} and by using a factor of 0.66 to correct from a Salpeter to a Kroupa IMF,

\begin{equation} \label{eq:sfr_LNUV}
    \frac{SFR_{NUV}}{M_{\odot}\ yr^{-1}} =
\frac{L_{NUV}}{2.14 \times 10^9}
 \times 0.66
\end{equation}

In Fig. \ref{fig:sfms} we plot the star-forming sequence (SFS) of the SAGAbg-morph galaxies with flux measurements made in the GALEX NUV band. The points represent fluxes with S/N$>4$ while the upside-down triangles show the upper limit of the measurements with S/N$<4$. In the left panel that shows the SFR plotted against the stellar mass, the points are colored by the $g-r$ color given in the SAGAbg catalog. The dotted and dash-dotted lines correspond to the best-fit slopes of the SFS measured in H$\alpha$ by \citet{KadoFong2024sagabg2} of galaxies belonging to the redshift ranges- $0.0 < z_{\rm spec} < 0.035$ and $0.07 < z_{\rm spec} < 0.105$ respectively. The median uncertainty of the SFR estimates of the measurements with S/N$>4$ is found out to be $\sim 0.1$ dex.

We find the SFS as traced by the GALEX measurements are in good agreement with the literature SFS. We also notice the gradient in $g-r$ color matches that of the SFR i.e., blue colors with increasing SFR. In the right panel we plot the specific star-formation rate (sSFR) in NUV that is the ratio $SFR/M_{\ast}$. The dash-dot line shows the value of ${\rm log}({\rm sSFR/yr}^{-1}) = -11.0$, which is employed as a threshold of quenched versus star-forming systems for the SAGA satellite sample in \citet{Geha2024saga4}.


\subsection{LS Image Processing} \label{sec:image_proc}

We use PHOTUTILS \citep{larry_bradley_2025_14889440} for processing the multi-band galaxy maps of the SAGAbg-morph sample. We are cognizant that the images in one or more of the bands have low signal-to-noise that motivates us to develop a unique pipeline for processing them. We run several trials using PHOTUTILS across a range of a few hundred galaxy maps, to identify failures and determine the best set of parameters that can reliably provide the segmentation maps of the SAGAbg-morph sources that are required for extracting the morphologies using STATMORPH. Throughout this process we assign flag variables ${\rm F}[b]$ for each of the sources in each of $b \in \{g,r,i,z\}$. We initialize these Boolean variables with ${\tt True}$ and assign them with ${\tt False}$ in case of failures at stages of the pipeline.

We determine the background, by masking objects above a $1\sigma$ threshold, having at least 10 pixels and connected on the sides as well as diagonally, i.e. 8-connectivity scheme. We prepare a mask and then use the \texttt{Background2d} class to prepare a low-level representation of the background using a $32\times32$ pixel box and a $\sigma=5$ pixel filter for smoothening the low-resolution map. Alongside the background subtracted galaxy map, we obtain the RMS map as well. Failures at this stage include the inability to detect objects the $1\sigma$ threshold and improper values in the inverse-variance maps, which are accordingly flagged using the ${\rm F}[b]$ variable.

\subsection{Source Detection \& Segmentation} \label{sec:segment}

\begin{figure*} 
\centering
\includegraphics[scale=0.38]{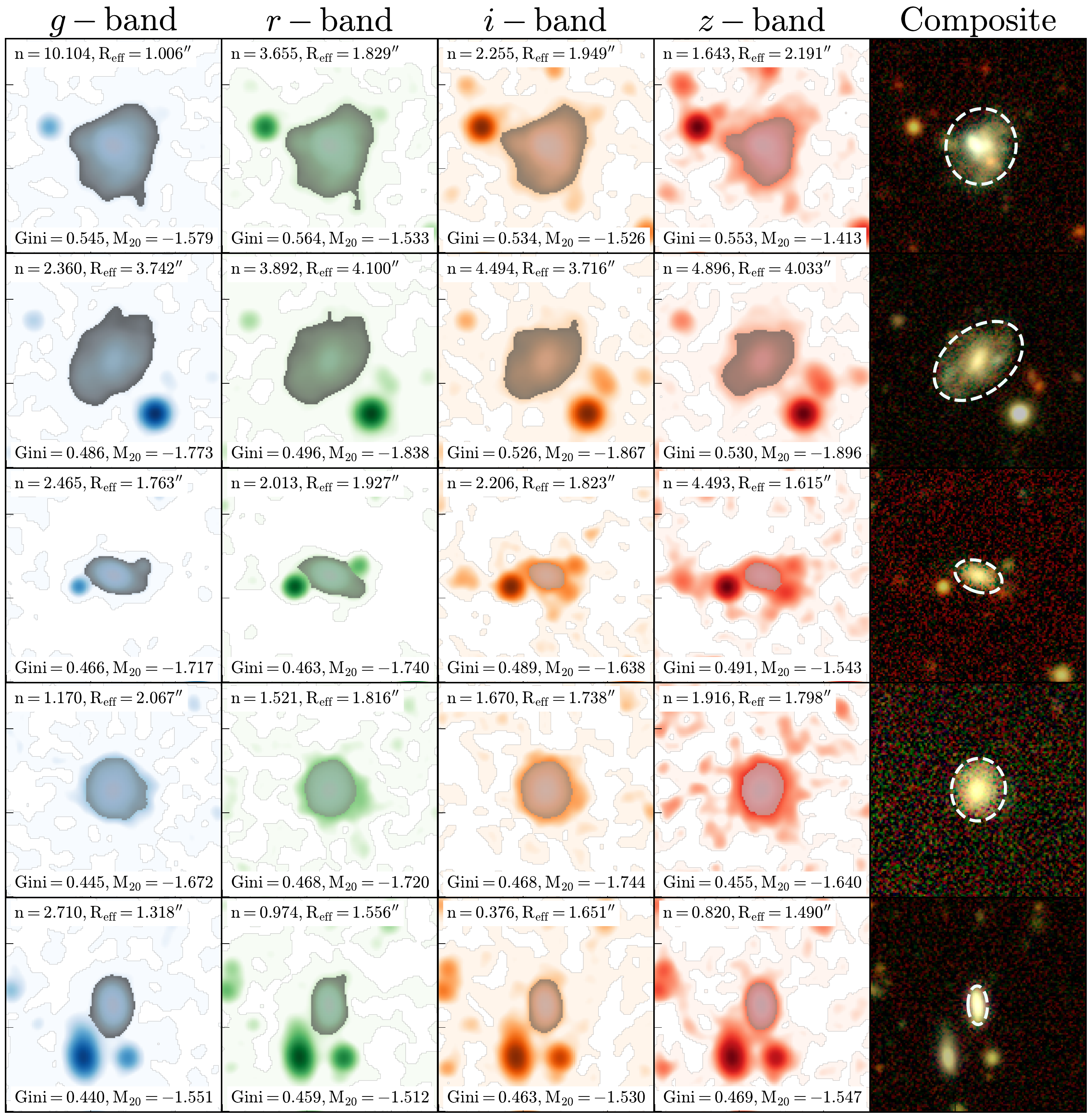}
\caption{Gallery of representative galaxy maps from the SAGAbg-morph sample showing $griz$ band imaging. Each row displays a single galaxy observed in (left to right), with a false-color composite image in the rightmost panel. The segmentation maps in each of the bands are plotted on the galaxy maps as gray regions. The ellipse outlined by the \textit{white dashed line} in the rightmost column corresponds to the Petrosian radius in the $r$-band.  The sample showcases a range of diverse sizes and morphologies that include galaxies of irregular, star-forming disk and spheroidal types. The presence of bright objects separated from the galaxy shows the success of the segmentation and deblending procedure employed in this work.  \label{fig:segmap_cutouts}}
\end{figure*}

We obtain the segmentation map from the multi-band, background subtracted galaxy maps. This is a non-trivial task on account of the image depth, crowded field, foreground stars and blending. Therefore we devised a systematic way to select the pixels associated with the target objects while minimizing contamination from surrounding sources. We optimize the parameters associated with the process through rigorously testing the segmentation map in relation to the multi-band galaxy maps. 

We start by convolving the maps with a PSF modeled using a Moffat function to smooth out pixel-to-pixel noise. We adopt the prescription in  \citet{Sazonova2025statmorph} where a Moffat PSF that approximates the effect of atmospheric seeing on the empirical PSF \citep{2001MNRAS.328..977T}, is chosen. We choose a FWHM that has a slightly larger size of $\sigma=3$ pixels and accordingly set the parameters for the Moffat PSF as $\beta=2$ and $\alpha = \frac{1}{2}(2^{1 / \beta} - 1)^{-1/2} {\rm FWHM} $. We then use the \texttt{SourceFinder} routine of PHOTUTILS to detect sources in the map 1$\sigma$ above the RMS map with at least 10 pixels, using 32 multi-thresholding levels and a deblending contrast of 0.0005. We find the target source as the one closest to the center. We then prepare segmentation maps in each of the bands $\mathbf{S}[b],b\in(g,r,i,z)$ by selecting the pixels that have been assigned to the target by \texttt{SourceFinder}. Alongside this we also prepare masks $\mathbf{M}[b],b\in(g,r,i,z)$, where we identify the pixels belonging to other bright sources in the map, with the aim of using this for the S{\'e}rsic fitting.


We combine the segmentation maps for each of the bands at the pixel-level using the following scheme:
\begin{equation} \label{eq:comp_segm}
 \mathbf{S}_{\rm comp} =   (\mathbf{S}[g] \wedge \mathbf{S}[r]) \wedge ((\mathbf{S}[i]\oplus   {\rm F}[i])\vee (\mathbf{S}[z]\oplus {\rm F}[z])).
\end{equation}
Here $\wedge,\vee,\oplus $ are the symbols for the logical AND, OR and XOR operators respectively. This scheme has been designed knowing that the SAGAbg sources are bright in the $gr$ bands and any background/foreground interloping object will be redder. With this scheme we are able to increase the weight of pixels in the $gr$ bands relative to $iz$ as well as deal with the sources without proper $i$-band imaging, i.e. by assigning ${\rm F}[i]={\tt False}$.

We also prepare a mask of background objects in the cutout excluding the segmented region. The purpose of this is that STATMORPH explicitly requires this to perform the S{\'e}rsic fits of the source galaxies.

\begin{equation} \label{eq:comp_mask}
 \mathbf{M}_{\rm comp} =   (\mathbf{M}[g] \vee \mathbf{M}[r]) \vee (\mathbf{M}[i]\vee \mathbf{M}[z]).
\end{equation}


In Fig. \ref{fig:segmap_cutouts} we display image cutouts of representative sources from the SAGAbg-morph sample in the $griz$ bands and composite color across the different columns from left to right. The $griz$ cutouts have been smoothed with a $\sigma=3$ filter to reduce pixel-to-pixel noise. The gray region in each of them corresponds to the segmentation map $\mathbf{S}[b]$ in the corresponding band $b\in(g,r,i,z)$. The composite color image has been prepared from the $grz$ band fluxes using the prescription of \citep{2004PASP..116..133L} and is shown in the extreme right column. The composite segmentation map $\mathbf{S}_{\rm comp}$ is shown in this column by highlighting the pixels that belong to it. The ellipse outlined by the white shaded line in the same column corresponds to the Petrosian radius in the $r$-band. The consequent measures of $Gini,M_{20}$ and the $n,R_{\rm eff}$ from the S{\'e}rsic fit in each of the four bands are also stated for the $griz$ cutouts. The set of morphologies, that also depict separate objects in close vicinity, shows the diversity of the SAGAbg-morph sample as well as the superior segmentation and deblending capabilities of our image processing pipeline.

\subsection{Non-Parametric Morphology Measures} \label{sec:statmorph}

We use STATMORPH \citep{2019Rodriguez-Gomez,Sazonova2025statmorph} to evaluate the non-parametric measures from the multiband imaging of the SAGAbg-morph sample. We run this routine on each of the $griz$ band galaxy maps along with the applications of the RMS maps and segmentation map we obtained in Sec. \ref{sec:segment}. We use the composite segmentation map (see Eq. \ref{eq:comp_segm}) on all the bands. The quantitative estimation of the morphologies include the Gini coefficient, $M_{20}$, CAS parameters and the shape asymmetry parameter. We define the parameters, and how we measure them from the data, in the remainder of this section.

The Petrosian radius \citep{1976ApJ...209L...1P} $R_p$ is a measure of the galaxy's extent and is essential for the calculations performed by STATMORPH. It is defined as the radius at which the surface-brightness $I(r)$ is some fraction $\eta$ times the mean surface-brightness $\langle I(r) \rangle$,

\begin{equation}
I(R_p) = \eta \langle I(R_p) \rangle,
\end{equation}

with $\eta=0.2$. As a result, $R_p$ is insensitive to S/N and surface-brightness limits. We use the Petrosian radius calculated using an elliptical aperture in STATMORPH where it is assumed that the center is at the point minimizing the asymmetry index. 

The Gini coefficient measures the homogeneity of the flux distribution in a galaxy map \citep{2003ApJ...588..218A,2004AJ....128..163L}. For example, a galaxy map with all of the flux in a single pixel will give a coefficient of 1 whereas for a perfectly even distribution of flux the coefficient will be 0. This coefficient is calculated in STATMORPH by first smoothing the map with a filter of size $\sigma = R_p/5$ to ensure smoothness of the new segmentation map being made. Using the mean surface-brightness at $R_p$ as a threshold the map is computed and the un-smoothed pixels within this is used to calculate $Gini$ as follows:

\begin{equation}
Gini = \frac{\sum_{i}^{n} (2i - n - 1)|X_i|}{n(n-1)|\bar{X}|}.
\end{equation}

Here $X_i$ is the flux of the $i^{\rm th}$ pixel sorted in ascending order by brightness and $|\bar{X}|$ is the mean of the absolute values of all fluxes associated with the target source. Absolute values are used here to mitigate noise from low $\langle {\rm S/N}\rangle \lesssim 3$ galaxy maps where flux values can often be negative \citep{2004AJ....128..163L}. 

The value of the $M_{20}$ statistic \citep{2004AJ....128..163L} is derived from the second-order moment of the flux distribution in a galaxy map. Since the flux values of pixels are weighted by their $r^2$, the spatial distribution of the brighter pixels is accounted for. The second moment of an $X$ set of pixels is defined as,

\begin{equation}
\mu_X = \sum_{i}^{X} I_i \left((x_i - x_0)^2 + (y_i - y_0)^2\right). 
\end{equation}

The center $(x_0, y_0)$ corresponds to the centroid of the pixels in the segmentation map. $M_{20}$ is defined as the logarithm of the ratio of second moment of the brightest pixels carrying 20\% of the galaxy's flux with respect to the total second moment of the pixels assigned to the galaxy with the segmentation map.

\begin{equation}
M_{20} = \log_{10}\left(\frac{\mu_{20}}{\mu_{100}}\right) 
\end{equation}

$M_{20}$ is measured by keeping the center of the galaxy as a free
parameter and circular symmetry is not imposed. This makes $M_{20}$ quite different from the Gini coefficient and $C_{CAS}$ and is sensitive to signatures of merging systems, e.g., multiple nuclei. For a very uniform distribution of a galaxy's light, the value of $M_{20}$ approaches zero whereas for a highly nuclear light distribution in a galaxy, $M_{20}$ will be reduced to a very negative value.

The concentration index $C_{CAS}$ is defined by the ratio of curve of growth radii containing 80\% and 20\% of the total flux of a galaxy within $1.5R_p$ respectively \citep{2000AJ....119.2645B,2003ApJS..147....1C},

\begin{equation}
C_{CAS} = 5 \log_{10} \left(\frac{R_{80}}{R_{20}}\right)  
\end{equation}

This parameter is simply proportional to the bulge strength of the galaxy, i.e., the more the light profile is concentrated, the larger is the value of $C_{CAS}$. Along with $C_{CAS}$ \citet{2003ApJS..147....1C} also define the the asymmetry and smoothness indices- $A_{CAS}$, $S_{CAS}$, that are measures of disturbance and clumpiness in the galaxy's light distribution. These are calculated by subtracting from the original galaxy image the image rotated by 180$^{\circ}$ and the image smoothed with a boxcar filter of width $\sigma=0.25R_p$ respectively,

\begin{equation}
    A_{CAS} = \frac{\sum_{i,j} \left|I_{ij} - I_{ij}^{180}\right|}{\sum_{i,j} \left|I_{ij}\right|} - A_{\text{bgr}}.
\end{equation}

\begin{equation}
    S_{CAS} = \frac{\sum_{i,j} \left(I_{ij} - I_{ij}^S\right)}{\sum_{i,j} I_{ij}} - S_{\text{bgr}}
\end{equation}

STATMORPH also performs a 2D S{\'e}rsic fit \citep{1963BAAA....6...41S} taking the mask of background objects $\mathbf{M}_{\rm comp}$ (see Eq. \ref{eq:comp_segm}) and the PSF model as inputs. Along with the S{\'e}rsic half-light radius $R_{1/2}^{\rm Sersic}$, the S{\'e}rsic index $n$ is returned that is a measure of the slope of the light profile, with a higher value of $n$ implying that most of the galaxy's light is in the central core. A low $n$ means a shallow light distribution. STATMORPH uses the native \texttt{astropy} fitting package to achieve this fit. We tested the S{\'e}rsic modeling using the \texttt{pysersic} package \citep{Pasha2023pysersic} on a small subset ($N=100$) of the SAGAbg-morph sample. However this proved ill-suited given the low S/N of the images. An analysis of S{\'e}rsic fits as undertaken in \citet{Asali2025saga6} for the full SAGAbg sample is therefore reserved for a future work.

\section{Results} \label{sec:results}

We have measured the $griz$ bands morphologies for the SAGAbg-morph sample. This multi-dimensional dataset enables us to explore both the systematics as well as the galactic properties leading to these observations.

\subsection{Measurements}

\begin{figure}
\centering
\includegraphics[scale=0.225]{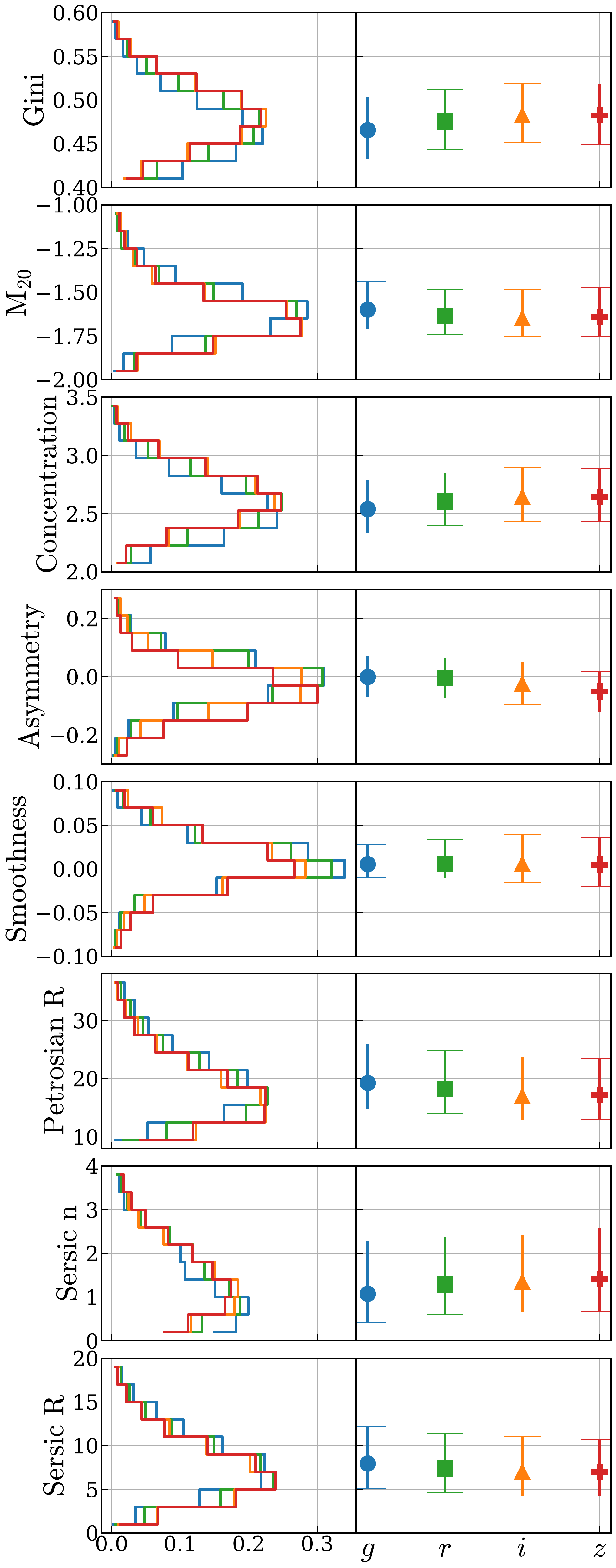}
\caption{The histograms and the median estimates for the relevant morphological measures of interest in the $griz$ bands are shown in the left and right columns. The estimates are made after imposing a quality based selection (see Sec. \ref{sec:diagnostics}). The error-bars on the median estimates reflect the 16-84$^{\rm th}$ percentiles of the distribution. We find that the different morphologies do not vary significantly across the bands and that both $A_{CAS},\ S_{CAS}$ have negative values that are nonphysical. \label{fig:NPhists}}
\end{figure}

\begin{figure}
\centering
\includegraphics[scale=0.35]{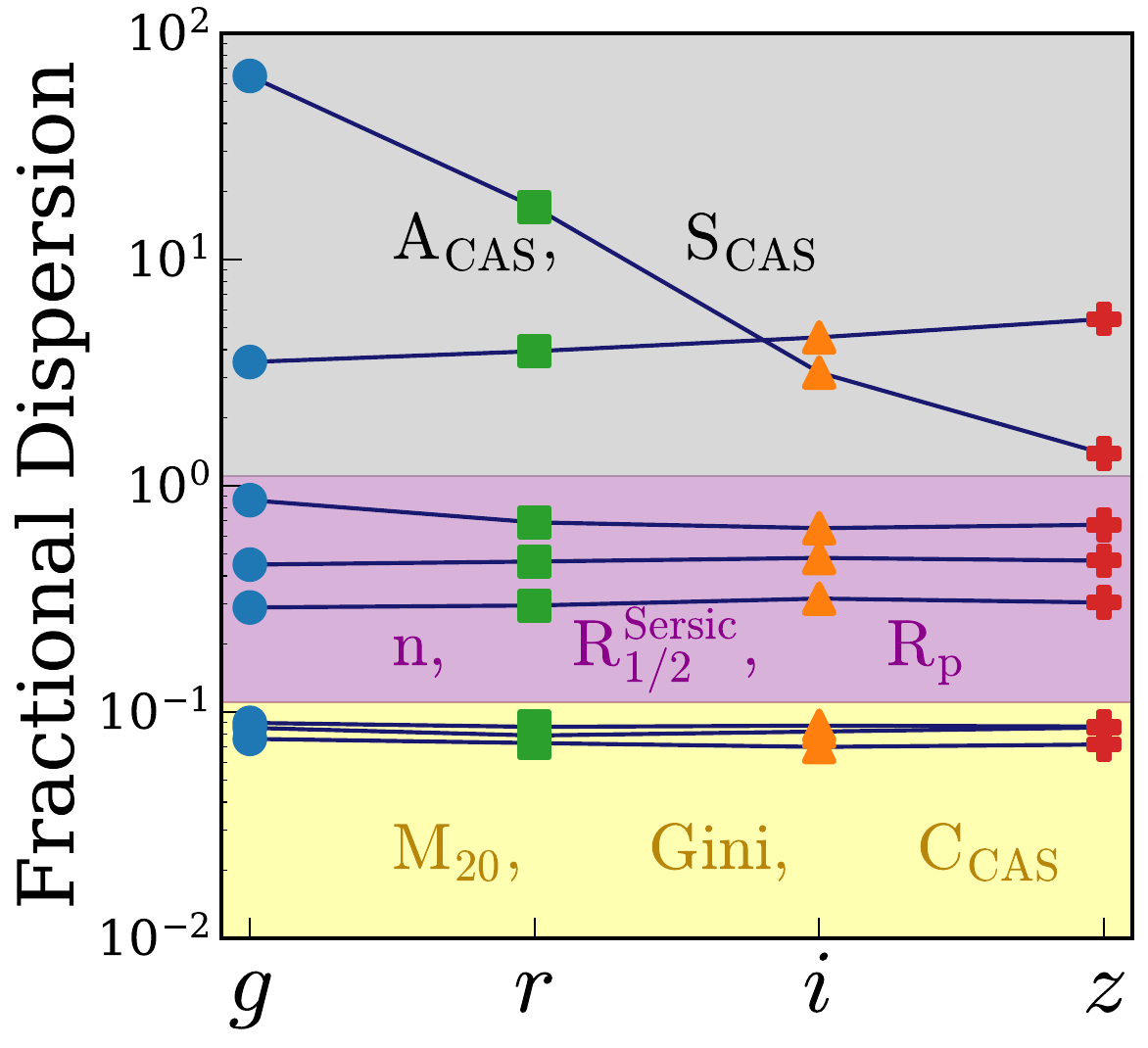}
\caption{The fractional dispersions of the morphological measures in the $griz$ bands. Each set of points connected by the \textit{solid lines} correspond to the measures and can be divided into three regions based upon the order-of-magnitudes of the fractional dispersions. The measures are appropriately annotated within their respective regions. The bulge prominence metrics- $M_{20},Gini,C_{CAS}$ occupying the yellow region being the most robust with this fraction being $<10\%$. The S{\'e}rsic model parameters, Petrosian radius (both in the purple region) $A_{CAS}$ and $S_{CAS}$ (both in the gray region) get progressively less reliable with larger values of dispersions. \label{fig:NPuncert}}
\end{figure}

\begin{table*}[h]
\centering
\caption{The measurements on the SAGAbg-morph sample that are made using STATMORPH. The estimate column shows the medians and their uncertainties that are plotted in Fig. \ref{fig:NPhists}. The $MI$ column displays the mutual information values that are shown in Fig. \ref{fig:MI_hist}. The $\rho_{M_{\ast}}$ and $\rho_{sSFR}$ columns show the Pearson correlation coefficients of the measurements with ${\rm log}(M_{\ast}/M_{\odot})$ and ${\rm log}({\rm sSFR/yr}^{-1})$ respectively. \label{tab:parameters}}
\begin{tabular}{l|l|ccccc}
\hline
Parameter Name & Description & Band &Estimate & $MI$ & $\rho_{M_{\ast}}$ & $\rho_{sSFR}$ \\
\hline
\multicolumn{5}{l}{Radius measurements} \\
\hline
$R_{p}$ & Petrosian radius &$g$ & $ 5.044 ^{- 1.163 }_{+ 1.761 }$ & 1.37 & 0.912 & -0.968 \\ 
(pixel)& & $r$ & $ 4.786 ^{- 1.119 }_{+ 1.711 }$ & - & 0.937 & -0.990 \\ 
& & $i$ & $ 4.465 ^{- 1.083 }_{+ 1.751 }$ & 1.52 & 0.952 & -0.961 \\ 
& & $z$ & $ 4.499 ^{- 1.103 }_{+ 1.64 }$ & 1.27 & 0.899 & -0.976 \\ 
\hline
$R_{1/2}^{\text{S{\'e}rsic}}$ & S{\'e}rsic half-light radius & $g$ & $ 2.082 ^{- 0.753 }_{+ 1.113 }$ & 1.38 & 0.943 & -0.952 \\ 
(pixel)& & $r$ & $ 1.932 ^{- 0.73 }_{+ 1.057 }$ & - & 0.876 & -0.985 \\ 
& & $i$ & $ 1.844 ^{- 0.73 }_{+ 1.04 }$ & 1.61 & 0.936 & -0.993 \\ 
& & $z$ & $ 1.826 ^{- 0.715 }_{+ 0.989 }$ & 1.26 & 0.740 & -0.765 \\ 
\hline
\multicolumn{5}{l}{Bulge strength measurements} \\
\hline
$n$ & Sérsic index &$g$ & $ 1.074 ^{- 0.648 }_{+ 1.207 }$ & 0.73 & 0.312 & -0.531 \\ 
& & $r$ & $ 1.291 ^{- 0.697 }_{+ 1.082 }$ & - & 0.965 & -0.797 \\ 
& & $i$ & $ 1.355 ^{- 0.695 }_{+ 1.066 }$ & 1.01 & 0.891 & -0.726 \\ 
& & $z$ & $ 1.423 ^{- 0.756 }_{+ 1.158 }$ & 0.7 & 0.617 & -0.867 \\ 
\hline
$C_{CAS}$ & CAS concentration index &$g$ & $ 2.538 ^{- 0.205 }_{+ 0.249 }$ & 1.0 & 0.430 & -0.999 \\ 
& & $r$ & $ 2.604 ^{- 0.204 }_{+ 0.244 }$ & - & 0.958 & -0.986 \\ 
& & $i$ & $ 2.646 ^{- 0.211 }_{+ 0.25 }$ & 1.14 & 0.995 & -0.977 \\ 
& & $z$ & $ 2.643 ^{- 0.209 }_{+ 0.247 }$ & 0.92 & 0.962 & -0.979 \\ 
\hline
$Gini$ & Gini coefficient & $g$ & $ 0.466 ^{- 0.033 }_{+ 0.038 }$ & 0.5 & -0.972 & 0.224 \\ 
& & $r$ & $ 0.475 ^{- 0.032 }_{+ 0.037 }$ & - &-0.978 & -0.691 \\ 
& & $i$ & $ 0.483 ^{- 0.032 }_{+ 0.036 }$ & 0.47 & -0.873 & -0.283 \\ 
& & $z$ & $ 0.482 ^{- 0.033 }_{+ 0.036 }$ & 0.39 & -0.959 & -0.756 \\
\hline
$M_{20}$ & Second-order moment of the & $g$ & $ -1.6 ^{- 0.112 }_{+ 0.161 }$ & 0.86 & -0.984 & 0.911 \\ 
& brightest 20\% of a galaxy’s pixels & $r$ & $ -1.637 ^{- 0.106 }_{+ 0.152 }$ & - & -0.999 & 0.930 \\ 
& & $i$ & $ -1.647 ^{- 0.107 }_{+ 0.163 }$ & 0.95 & -0.996 & 0.909  \\ 
& & $z$ & $ -1.642 ^{- 0.11 }_{+ 0.17 }$ & 0.77 & -0.959 & 0.911 \\ 
\hline
\multicolumn{5}{l}{Other measurements} \\
\hline
$A_{CAS}$ & CAS asymmetry index & $g$ & $ -0.001 ^{- 0.068 }_{+ 0.072 }$ & 0.47 & 0.906 & 0.999 \\ 
& & $r$ & $ -0.004 ^{- 0.069 }_{+ 0.068 }$ & - & 0.990 & 0.954 \\ 
& & $i$ & $ -0.023 ^{- 0.073 }_{+ 0.074 }$ & 0.42 & 0.977 & 0.998 \\ 
& & $z$ & $ -0.05 ^{- 0.072 }_{+ 0.068 }$ & 0.3 & 0.957 & 0.884 \\ 
\hline
$S_{CAS}$ & CAS smoothness index\
& $g$ & $ 0.005 ^{- 0.015 }_{+ 0.022 }$ &  0.24 & 0.976 & -0.671 \\ 
& & $r$ & $ 0.006 ^{- 0.016 }_{+ 0.028 }$ & - & -0.902 & 0.707 \\ 
& & $i$ & $ 0.006 ^{- 0.022 }_{+ 0.034 }$ & 0.28 & 0.971 & 0.847 \\ 
& & $z$ & $ 0.005 ^{- 0.025 }_{+ 0.031 }$ & 0.19 & -0.997 & -0.091 \\
\hline
\end{tabular}
\end{table*}

In Fig. \ref{fig:NPhists} we plot the histograms of the STATMORPH morphologies in the left column while the right column shows the median with the error-bar signifying the 16-84$^{\rm th}$ percentiles of the distribution. These estimates are computed from the raw distributions after applying the quality cuts that we discuss in Sec \ref{sec:diagnostics}. We show the fractional dispersions of the different metrics in Fig. \ref{fig:NPuncert}. The dispersions are derived from the $1\sigma$ confidence interval of the distributions described above. In Table \ref{tab:parameters} we list all of the morphology measures made using STATMORPH that are relevant to our analysis along with their medians estimates and their confidence intervals. 

In Fig. \ref{fig:NPuncert} we find that the morphologies show no significant variation across the different bands within the limits of their uncertainties. We also find that the Gini indices, $M_{20}$ statistics and the $C_{CAS}$ parameters have values within the appropriate dynamic ranges \citep{2019Rodriguez-Gomez}. However, the $A_{CAS}$ and $S_{CAS}$ indices show negative values that are nonphysical. Furthermore, the fractional uncertainties for $Gini, M_{20}$ and $C_{CAS}$ are $<10\%$ for all the bands whereas for the Petrosian and S{\'e}rsic radii measurements and the S{\'e}rsic indices it is $\sim 10-100\%$. The fraction dispersions serve as rule-of-thumb indicator of the quality of our measurements, since values $\gtrsim 100\%$ imply noise is dominant and therefore cannot be reliably analyzed. This analysis shows that only the bulge strength measurments- $Gini, M_{20}, C_{CAS}$ are robustly estimated with STATMORPH given the low S/N imaging data that we used. In the following sections \ref{sec:diagnostics} and \ref{sec:information} we explore the effect of noise and resolution further.

\subsection{Diagnostics} \label{sec:diagnostics}

\begin{figure*}
\centering
\includegraphics[scale=0.275]{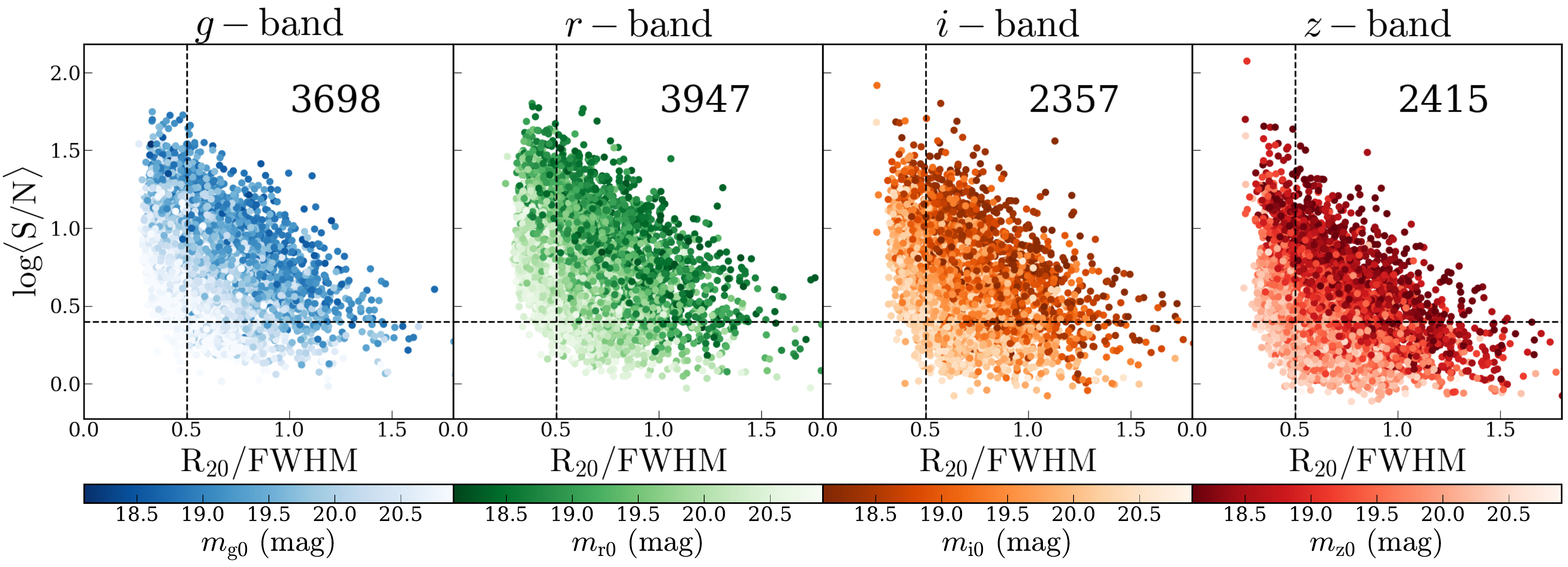}
\caption{Distribution of SAGAbg-morph sample in the signal-to-noise ratio (S/N) versus $R_{20}/$FWHM plane across $griz$ bands. Each panel shows the relationship between logarithmic $\langle \text{S/N}\rangle$ and the parameter $R_{20}/$FWHM (the ratio of the radius containing 20\% of the total flux to the PSF FWHM) for galaxies observed in the respective photometric band. Points are color-coded by apparent magnitude in each band, ranging from bright (dark colors, $\sim 18.5$ mag) to faint (light colors, $\sim 20.5$ mag). The number of galaxies in each band after the quality cut is indicated in the upper portion of each panel. Dashed lines mark the selection criteria at $R_{20}/{\rm FWHM} > 0.5$ (vertical) and $\langle \text{S/N}\rangle > 2.5$ (horizontal), defining the galaxies in sample with reliable STATMORPH outputs. The distributions show that fainter galaxies exhibit lower S/N and preferentially lower $R_{20}/$FWHM values, with the $r$-band showing the most robust measurements (3947 galaxies) \label{fig:S/Nfrac}}
\end{figure*}

It is necessary that we determine the quality of the STATMORPH calculations before we analyze and interpret the results. For this purpose we study the noise measurements and quality flags as returned by the routine \citep{2019Rodriguez-Gomez}. The noise is quantified by $\langle \text{S/N}\rangle$, the average ratio of the image map to the RMS map over the pixels that lie within an elliptical aperture defined by the Petrosian radius. STATMORPH also returns the \texttt{flag} and \texttt{flag\_sersic} to alert the user about issues arising during calculation of the non-parametric measures and the S{\'e}rsic model respectively. It is advised in \citet{2019Rodriguez-Gomez} that $\langle \text{S/N}\rangle > 2.5$, ${\tt flag}=0$ and ${\tt flag\_sersic}=0$ be enforced for reliable measurements. It is also noted that $R_{20}$, which is the radius containing 20\% of a galaxy’s flux, should be larger than half of the PSF FWHM.

In Fig. \ref{fig:S/Nfrac} we plot the logarithmic S/N per pixel and the $R_{20}/$FWHM parameter for the galaxies in SAGAbg-morph that were subject to STATMORPH measurements. While the S/N per pixel is a proxy for depth, the parameter on the horizontal axis corresponds to resolution. These correspond to the two leading effects behind the systematics associated with morphology measurements \citep{Sazonova2025statmorph}. The four panels show the $griz$ bands with each point colored according to the apparent magnitude (from the TRACTOR catalog) corrected for reddening in the respective band. The vertical and horizontal dashed lines represent the limits of our selection at $R_{20}/{\rm FWHM} = 0.5$ and $\langle \text{S/N}\rangle > 2.5$ respectively.  After accounting for the quality flags we find that 3698, 3947, 2357 and 2415 sources in each of the $griz$ bands have reliable STATMORPH measurements. We use these sub-sample of measurements that pass the quality checks throughout the rest of the analysis in this work. We find that the median $\langle \text{S/N}\rangle$ decreases from the $g$ to $z$ bands while the median $R_{20}/$FWHM parameter increases. Due to the competing trends the $r$-band produces the most reliable measurements as shown by the largest number of sources in this band.

\subsection{Mutual Information} \label{sec:information}

\begin{figure}
\centering
\includegraphics[scale=0.4]{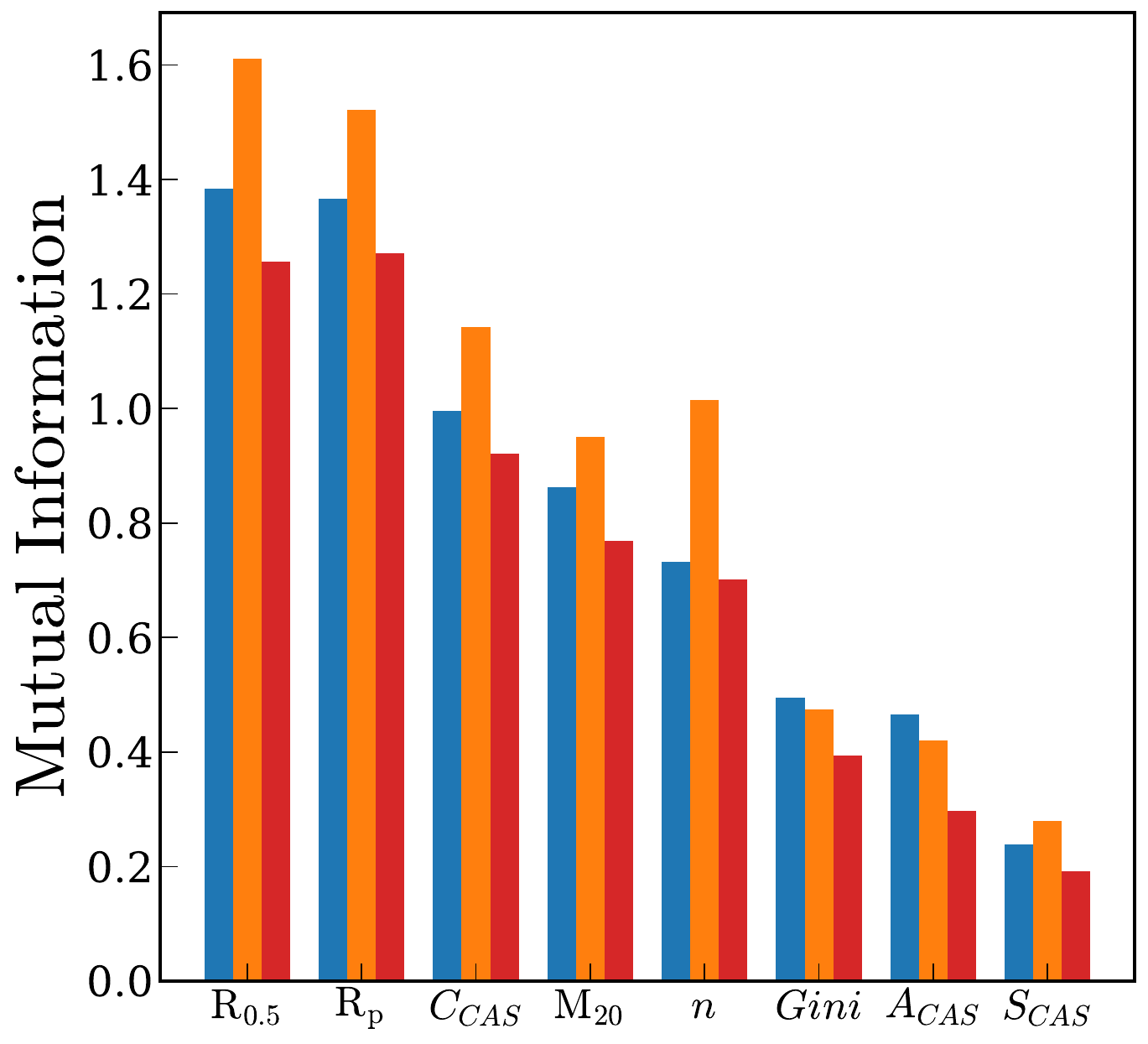}
\caption{ The mutual information $MI$ between the $r$-band morphology measure and the respective measures in the $giz$ bands are show using the \textit{blue, orange, red} colored bar plots. Aside from the radius measurements, the STATMORPH outputs tracing the bulge strength- $C_{CAS}$, S{\'e}rsic $n$ and $M_{20}$ are more robust compared to $A_{CAS}$ and $S_{CAS}$. \label{fig:MI_hist} }
\end{figure}


The large dataset of STATMORPH measurements gives us the ability to explore relationships among the bands as well as non-parametric and parametric measures. A very effective way to look at the utility of studying multi-band morphologies is to use an information theory based approach. The mutual information score has been used to study the importance of morphological measures with respect to galaxy properties \citep{2021ApJ...923..205Y}. However in our work we use this score to gauge the usefulness of the STATMORPH output given the multi-band data rather than explore their connection to galaxy properties. We calculate the score $MI$ between the morphology measures in the $r$-band with those in the $giz$ bands. This tells us about the statistical dependence between the measures from different bands. Lack of correlation along with scatter tends to produce a low $MI$ score. Here we assume that the distributions of the STATMORPH morphologies are similar across the bands, i.e., any degradation in the information content of a band is due to measurement noise. We choose $r$-band measures as the fiducial band given its superior measurement quality. In Fig. \ref{fig:MI_hist} we plot our results using blue, orange, and red bars representing the $giz$ bands. These are arranged in descending order of $MI$ from left to right.

We find that the radii measurements in terms of the half-light $R_{1/2}^{\text{S{\'e}rsic}}$ and Petrosian $R_p$ exhibit the highest $MI$ across all the bands. These are followed by the bulge strength measurements- $C_{CAS}$, $M_{20}$ and S{\'e}rsic $n$ in the middle range of $MI$. Finally, the disturbance measurements that include $A_{CAS}$ and $S_{CAS}$ have the lowest $MI$ in all bands. Even though $Gini$ corresponds to bulge strength, its $MI$ score is relatively lower and comparable to those of the disturbance measurements. We also note that $MI$ is highest between $r$ and $i$ measures and decreases between $r$ and $g$, with $MI$ being the lowest between $r$ and $z$.

Our inferences are strongly in agreement with the studies where the dependence of the morphology indices on RMS noise and resolution are investigated \citep{2004AJ....128..163L,2011MNRAS.416.2401H,Sazonova2025statmorph}. While the radius measures- $R_p$ and $R_{1/2}^{\text{S{\'e}rsic}}$ show robustness with respect to depth and resolution at different redshifts, the bulge strength parameters of Gini, $M_{20}$, $C_{CAS}$ are sensitive to resolution. This is because as the PSF FWHM increases this leads to the spread of the light from the galaxy's nucleus over a large area, thereby decreasing bulge prominence \citep{Sazonova2025statmorph}. On the other hand, measures like $A_{CAS}$ and $S_{CAS}$ are highly sensitive to S/N. Their low $MI$ scores are also linked to the presence of nonphysical values of $A_{CAS}$ and $S_{CAS}$. Based on this analysis as well as the magnitudes of the fractional dispersions in Fig. \ref{fig:NPuncert}, we consider our disturbance measurements to be distorted and choose to make the the bulge strength metrics ($Gini, M_{20}, C_{CAS}$) the main focus of the rest of our work.

\subsection{Trends with Stellar Mass and sSFR} \label{sec:trends}

\begin{figure}
\centering
\includegraphics[scale=0.4]{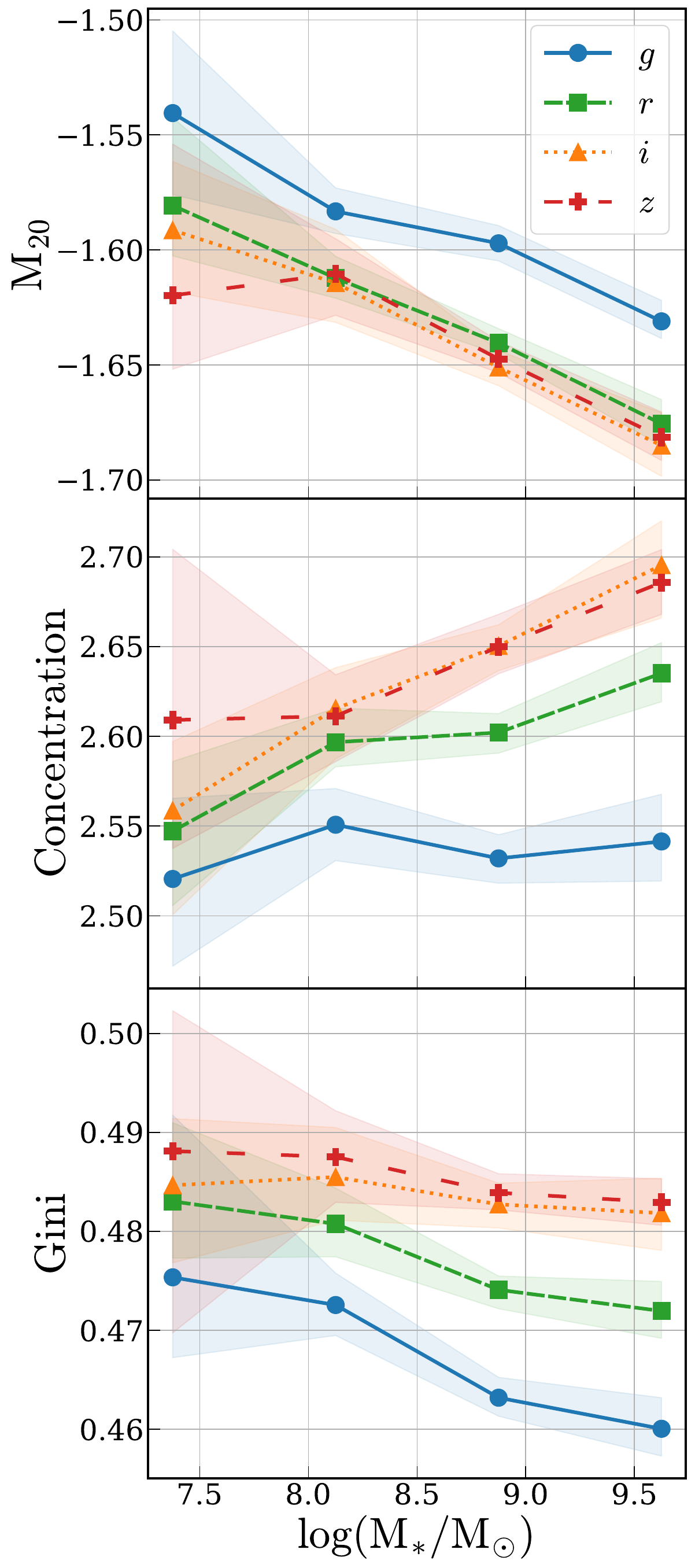}
\caption{Stellar mass dependence of morphological parameters in the SAGAbg-morph sample after imposing the quality cuts. Median trends of Gini index (\textit{top}), $M_{20}$ (\textit{center}), concentration $C_{CAS}$ (\textit{bottom}), as a function of stellar mass ${\rm log}(M_{\ast}/M_{\odot})$ for galaxies in $g$ (blue circles), $r$ (green squares), $i$ (orange triangles), and $z$ (red crosses) bands. The median estimates are evaluated from bootstrap distributions with the shaded regions indicating the $3\sigma$ confidence intervals. $M_{20}$ both become more negative and $C_{CAS}$ increases (indicating more concentrated light distributions) with increasing stellar mass. The systematic trends are qualitatively similar across all bands, but the $g$-band indicate an even light distribution, e.g., with higher values of $M_{20}$. \label{fig:NPSplineM}}
\end{figure}

\begin{figure}
\centering
\includegraphics[scale=0.4]{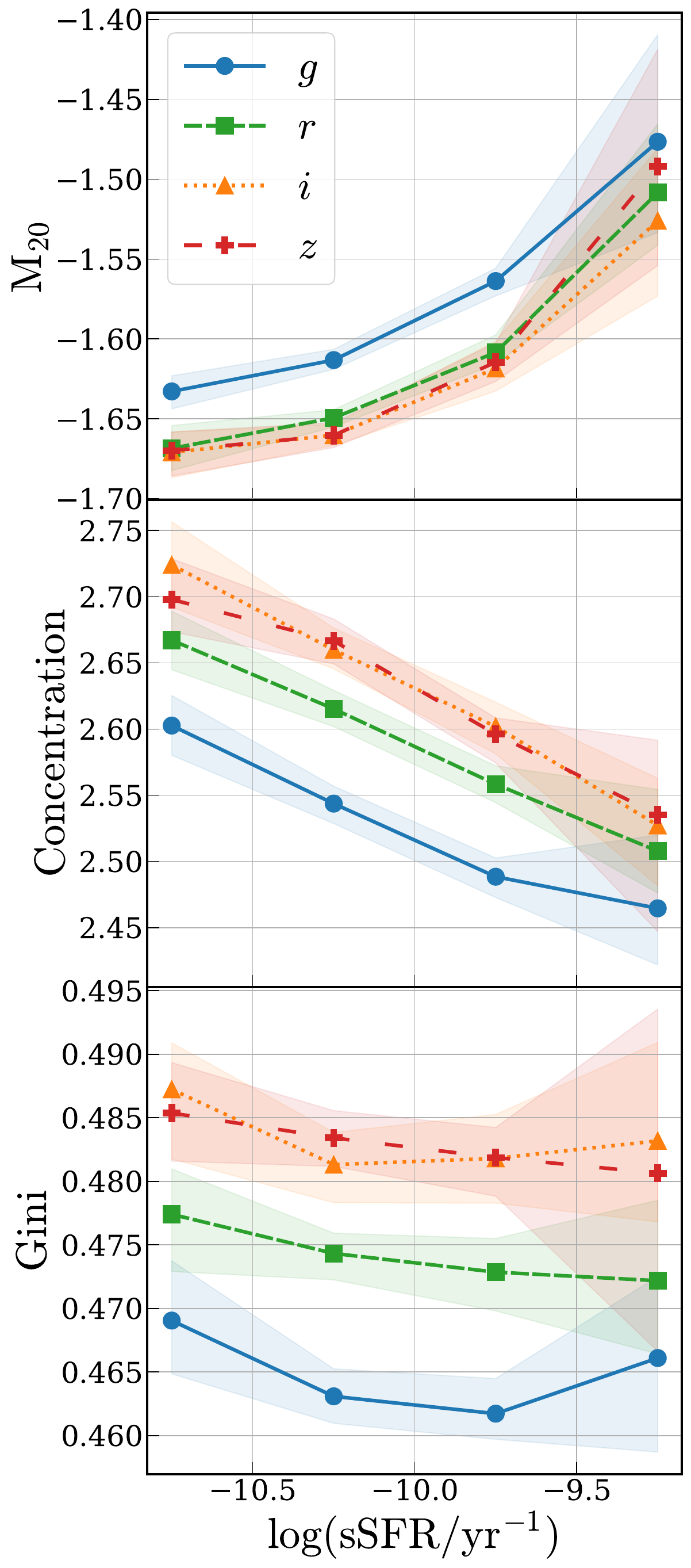}
\caption{sSFR dependence in the SAGAbg-morph sample after imposing the quality cuts. Median trends of Gini index (\textit{top}), $M_{20}$ (\textit{center}), concentration $C_{CAS}$ (\textit{bottom}), as a function of the specific star-formation rate (sSFR) ${\rm log}({\rm sSFR/yr}^{-1})$ for galaxies in $g$ (blue circles), $r$ (green squares), $i$ (orange triangles), and $z$ (red crosses) bands. The median estimates are evaluated from bootstrap distributions with the shaded regions indicating the $3\sigma$ confidence intervals. $M_{20}$ becomes less negative and $C_{CAS}$ decreases (indicating a more uniform light distributions) with increasing sSFR. Similar to Fig. \ref{fig:NPSplineM} we find that the $g$-band metrics imply light distributions that are less concentrated. \label{fig:NPSplineSSFR} }
\end{figure}

Here we investigate trends between the bulge strength measurements with the stellar masses and specific star-formation rates (sSFR) of the SAGAbg-morph sample. The stellar mass ${\rm log}(M_{\ast}/M_{\odot})$ is obtained from the SAGAbg catalog (see Eq. \ref{eq:stellar_mass} in Sec \ref{sec:SAGAbg}) whereas we use the sSFR that were determined using GALEX NUV flux measurements in Sec. \ref{sec:galex} (see Eq. \ref{eq:sfr_LNUV}). In the latter analysis, we only use all the GALEX flux measurements, including those with upper bounds. We compute the bootstrap distributions ($N=2500$) of the median statistics in each of the linearly spaced 0.75 dex bins in the interval of $7<{\rm log}(M_{\ast}/M_{\odot})<10$  and 0.5 dex bins of $-11<{\rm log}({\rm sSFR/yr}^{-1})<-9$ respectively. We then calculate the mean of the median distributions along with their $3\sigma$ confidence intervals. In Fig. \ref{fig:NPSplineM} and \ref{fig:NPSplineSSFR} we plot the bootstrap derived median trends in the $griz$ bands with the shaded bands representing the appropriate confidence intervals. 


In Fig. \ref{fig:NPSplineM} we find that  $M_{20}$ and $C_{CAS}$ each show declining and rising trends with respect to stellar mass across $griz$. This means that the galaxy's light profile is more concentrated at higher masses, implying a more prominent bulge. Among the bands, $M_{20}$ in the $g$-band is the largest at all masses while $C_{CAS}$ for the same is the lowest. The light in the smallest wavelength, i.e., $g$-band, is the most evenly distributed. If extinction is not biasing these measurements across the bands, this implies that younger stars/active star-formation as traced by the $g$-band flux, occurs more homogeneously across the extent of the galaxy. We note that $Gini$ digresses as declining trends are found in the $gr$ bands while no dependence on stellar mass is found in the $iz$ bands. A decreasing $Gini$ is discrepant with an increase in the bulge prominence. Since this coefficient is sensitive to both resolution and noise \citep{Sazonova2025statmorph} as compared to $M_{20}$ and $C_{CAS}$ we argue that the divergence among the bands is due to these systematics. The corrections to $Gini$ of magnitude $\sim 0.01-0.02$ that we calculate in Appendix \ref{sec:correctGM20} are entirely sufficient to explain this discrepancy.

We notice that in Fig. \ref{fig:NPSplineSSFR}, $M_{20}$ and $C_{CAS}$ shows rising and declining trends respectively with increasing sSFR, whereas the Gini index show no significant variation with sSFR. However, this again shows that with increasing sSFR the galaxy's light distribution becomes less concentrated, i.e., signifies weaker bulges. This is expected since sSFR is inversely proportional to stellar mass (right panel of Fig. \ref{fig:sfms}) and is therefore in line with the systematic trends w.r.t. stellar mass. More so, the largest values of $M_{20}$ and smallest values of $C_{CAS}$ among all the bands occur in the $g$-band at all values of sSFR. 

\subsection{Gini-M20 Distribution} \label{sec:GiniM20_space}

\begin{figure}
\centering
\includegraphics[scale=0.26]{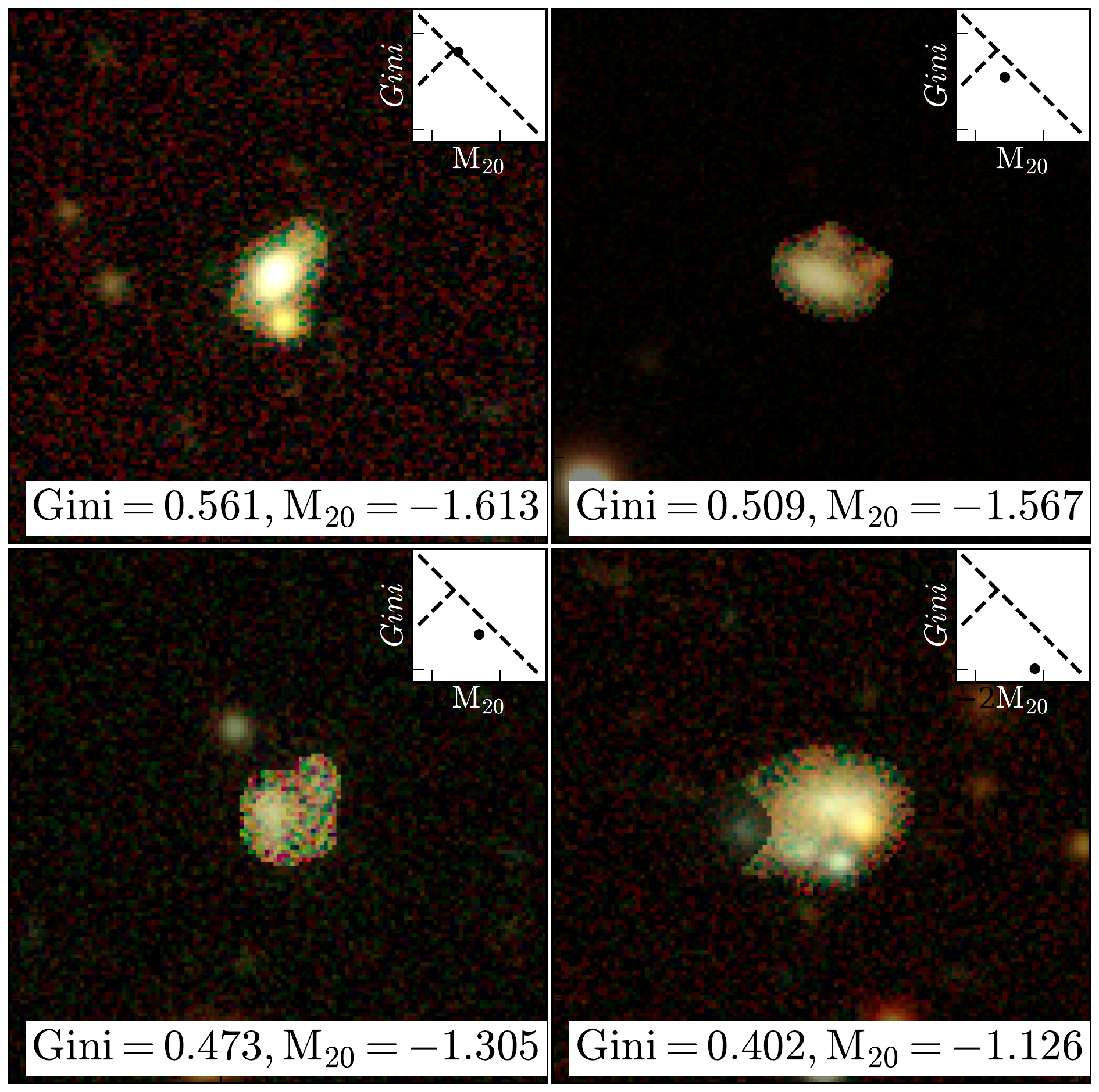}
\caption{Images of four representative galaxies with decreasing values of $Gini$ and increasing $M_{20}$ through the top left, top right, bottom left and bottom right panels. The insets in the corners show the position of the galaxies in the $Gini-M_{20}$ space and we find that the galaxies have progressively less concentrated light profiles, i.e., more disk dominated morphologies.\label{fig:GM20_2column}}
\end{figure}

\begin{figure*}
\centering
\includegraphics[scale=0.4]{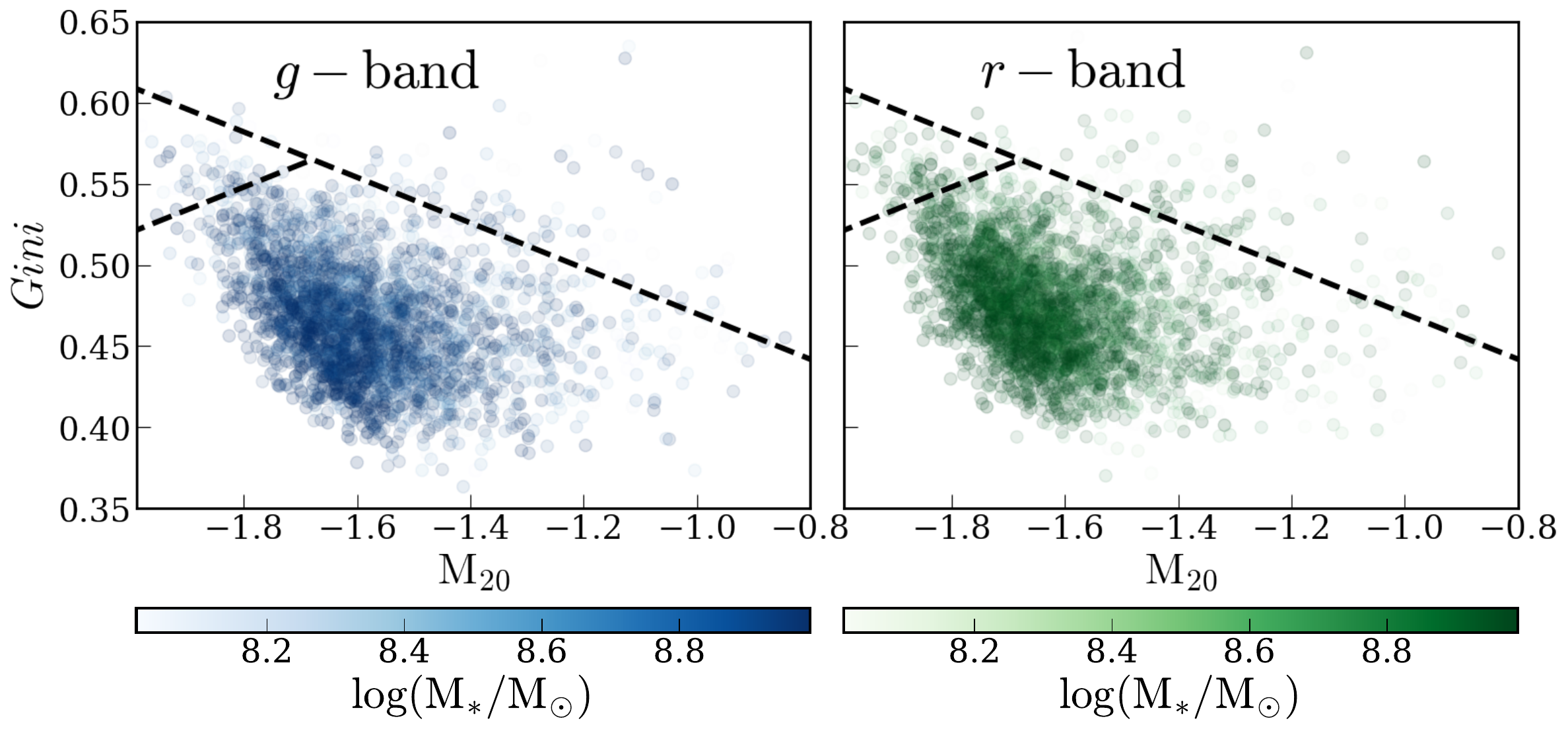}
\caption{$Gini-M_{20}$ space distributions for the SAGAbg-morph sample. \textit{Left}: $g$-band measurements. \textit{Right}: $r$-band measurements. Points are color-coded by stellar mass. \textit{Dashed} lines delineate the empirical division between merger/irregular systems (upper half) and relaxed disk/spheroid galaxies (lower half). The sequence as seen in the dense cluster of points persists across both bands. Their loci in this space correspond to Sb/Sc/Ir morphologies, i.e., disk or irregular galaxies.  \label{fig:GM20_scatter}}
\end{figure*}

The joint distribution of the Gini coefficient and the $M_{20}$ statistic can be empirically linked to the visual classification of massive galaxies \citep{2004AJ....128..163L,2008ApJ...672..177L}. The E/S0/Sa classes of bulge-dominated galaxies have high Gini indices and low $M_{20}$ values. The Sb/Sc/Ir classes correspond to galaxies with star-forming disks resulting in low Gini indices and high $M_{20}$ values. 

\citet{2008ApJ...672..177L} outlines a morphological classification scheme based on the $Gini-M_{20}$ distribution of massive galaxies. The galaxies in the region of the $Gini-M_{20}$ space corresponding to,
\begin{equation} \label{eq:merger_line1}
Gini \geq -0.14 M_{20} + 0.33\ {\rm and}\ Gini > 0.14 M_{20} + 0.80    
\end{equation}
are populated by the E/S0/Sa galaxies with strong bulges and early-type morphologies. As opposed to this, the region defined by, 
\begin{equation} \label{eq:merger_line2}
Gini \leq -0.14 M_{20} + 0.33\ {\rm and}\ Gini < 0.14 M_{20} + 0.80    
\end{equation}
corresponds to the Sb/Sc/Ir galaxies with weak bulges, star-forming disks or irregular morphologies. Lastly,
\begin{equation} \label{eq:merger_line3}
Gini \geq -0.14 M_{20} + 0.33    
\end{equation}
usually contains galaxies undergoing major-mergers as shown by their disturbed morphologies.

In Fig. \ref{fig:GM20_scatter} we show the composite color images of four galaxies that are selected in different regions of the $Gini-M_{20}$ space. The values of the Gini coefficient decrease and $M_{20}$ increase going from the top left, top right, bottom left and bottom right panels of the figure. Here we find that as we move from the left half to the bottom half of the space along the diagonal, the morphologies of the SAGAbg-morph galaxies have less concentrated light distributions as we would expect according to the \citet{2008ApJ...672..177L} scheme. 

In Fig. \ref{fig:GM20_scatter} we plot the scatter of the full sample in the $Gini-M_{20}$ space for the $g$ and $r$-bands. The points are color coded with ${\rm log}(M_{\ast}/M_{\odot})$ such that the darker points are the more massive galaxies.  We choose to only use these bands for the analysis because these have relatively higher S/N compared to the $iz$ bands. We find that the SAGAbg-morph galaxies occupy the region of the space corresponding to Sb/Sc/Ir galaxies in Eq. \ref{eq:merger_line2} in the two bands. The distribution of points in the $Gini-M_{20}$ spaces overall resemble a sequence, whose location corresponds to disky galaxies with weak bulges, and is in line with our expectations for such low-mass star forming galaxies. We further notice that the lower left edge of the sequence is populated by more massive galaxies. This shows that even with the sequence there exists a gradient in galaxy properties. This motivates us to analyze the $Gini-M_{20}$ joint distributions further by modeling these as bivariate Gaussian functions to study the variations with the physical properties of the galaxies. The details of the model and the fitting procedure are outlined in Appendix \ref{sec:2dgaussian}.

\begin{figure*}
\centering
\includegraphics[scale=0.35]{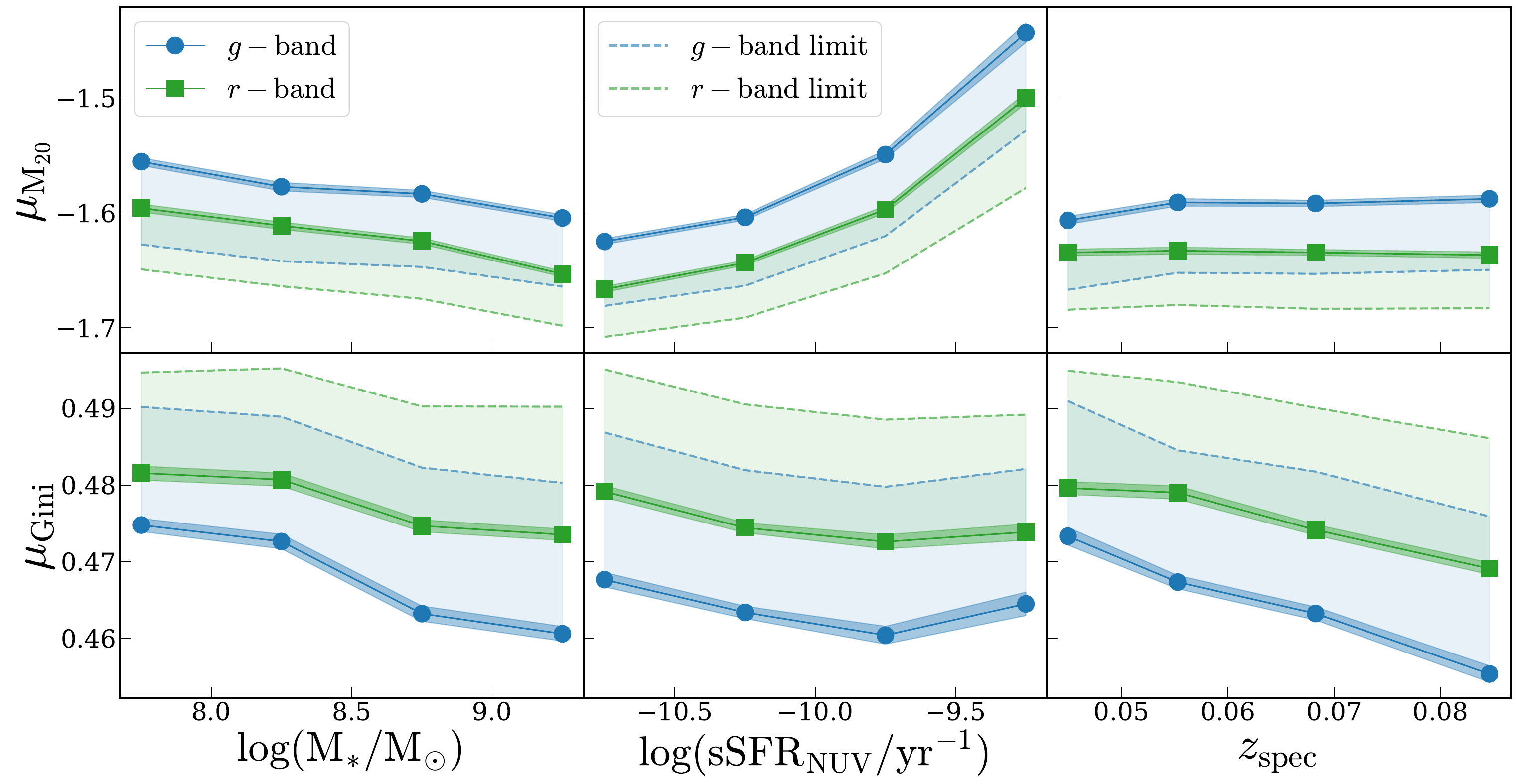}
\caption{Shift of the SAGAbg-morph sequence, as parameterized by a 2D Gaussian distribution, in $Gini-M_{20}$ space across stellar mass, sSFR and redshift. \textit{Top row}: Best-fit mean $M_{20}$ ($\mu_{M_{20}}$) versus $M_{\ast}$ (\textit{left}), sSFR (\textit{center}), and $z_{\rm spec}$ (\textit{right}). \textit{Bottom row}: Best-fit mean $Gini$ ($\mu_{\rm Gini}$) versus the same parameters. \textit{Blue circles} and \textit{green squares} represent $g$ and $r$-band measurements respectively, with shaded regions indicating $3\sigma$ confidence intervals. $M_{20}$ becomes more negative with increasing $M_{\ast}$ and decreasing sSFR, indicating that more massive and quiescent galaxies have more centrally concentrated light distributions. Both parameters show mild evolution with redshift, with systematic offsets between $gr$ bands reflecting wavelength-dependent morphological measurements. The systematic trends demonstrate that galaxy structure correlates more strongly with star formation activity and mass than with cosmic time over the $z_{\rm spec}<0.1$ range. \label{fig:GM20_redshift}}
\end{figure*}

We now investigate the dependence of the $Gini-M_{20}$ distribution with $M_{\ast}$, sSFR and $z_{\rm spec}$. We do this by selecting four linearly spaced bins of each of ${\rm log}(M_{\ast}/M_{\odot})\in(7.5,9.5)$, ${\rm log}({\rm sSFR/yr}^{-1})\in(-11.0,-9.0)$ and four logarithmically spaced bins of $z_{\rm spec}\in(0.004,0.1)$. We select the galaxies in the SAGAbg-morph sample with the corresponding stellar mass/sSFR/redshift in a particular bin, including all GALEX measurements- both the reliable detections as well as the upper-limits. We then fit bivariate Gaussian functions to their distribution in $Gini-M_{20}$ space using \texttt{lmfit} by following the procedure detailed in Appendix \ref{sec:2dgaussian}. Among the parameters of the bivariate Gaussians, we focus on the centroids- $\mu_{M_{20}},\mu_{\rm Gini}$ since we are primarily interested in the shift of this sequence with changes in the galaxy properties. 


In Fig. \ref{fig:GM20_redshift} we plot the best-fit values of the centroids in bins of $M_{\ast}$, sSFR and $z_{\rm spec}$. The dark shaded regions represent the $3\sigma$ uncertainties as derived from the covariance matrix given by \texttt{lmfit}. The light shaded region bordered by the dashed line represents the extents of corrections on $Gini$ and $M_{20}$, as calculated using the prescription of \citet{Sazonova2025statmorph} and detailed further in App. \ref{sec:correctGM20}. We find both $\mu_{\rm Gini}$ and $\mu_{M_{20}}$ in both bands show monotonically decreasing trends with stellar mass, while $\mu_{M_{20}}$ rises with sSFR and $\mu_{\rm Gini}$ falls with sSFR. This is purely consistent with the median trends we found in Sec. \ref{sec:trends}. With $z_{\rm spec}$ however we find flat and decreasing trends for $\mu_{M_{20}}$ and $\mu_{\rm Gini}$ respectively.

\citet{Sazonova2025statmorph} highlight that both $Gini$ and $M_{20}$ measures are sensitive to resolution with the former having a weak dependence on S/N as well. These effects tend to blur out the galaxy's light such that the measured $Gini$ or $M_{20}$ erroneously implies a less concentrated light distribution. The work provides prescriptions to model the correction for these measures using symbolic regression. We calculate these corrections and find that while $Gini$ is under-estimated by $\sim 0.015$ and $M_{20}$ is over-estimated by $\sim 0.05$. These corrections alone cannot explain the systematic trends of $\mu_{M_{20}}$, particularly the one against sSFR. However the gently declining trends of $\mu_{\rm Gini}$ across $M_{\ast}$, sSFR and $z_{\rm spec}$ might vanish in light of these corrections. Although the systematics distort the $Gini-M_{20}$ distribution towards the lower-right corner, i.e., corresponding to disks with flatter light profiles, we reason that the physical shifts in our distribution due to variation in $M_{\ast}$ and sSFR cannot be simply attributed to such distortions alone.

We interpret the shifts in the sequence in $Gini-M_{20}$ across bins of the galaxy's physical properties as a reflection of the trends we observed in Sec. \ref{sec:trends}. Namely the increase in $\mu_{M_{20}}$ with increasing $M_{\ast}$ and decreasing sSFR point to the SAGAbg-morph galaxies having progressively more concentrated light profiles. The constant values of $\mu_{M_{20}}$ with respect to $z_{\rm spec}$ hint at no redshift evolution taking place. We claim that the strong variation in $M_{20}$, particularly with respect to sSFR, is a signature of transition from purely disk dominated to stronger bulges at this sSFR scale.

\subsection{Evolution of the Star Forming Sequence} \label{sec:sfms_ev}

\begin{figure*}[!ht]
\centering
\includegraphics[scale=0.275]{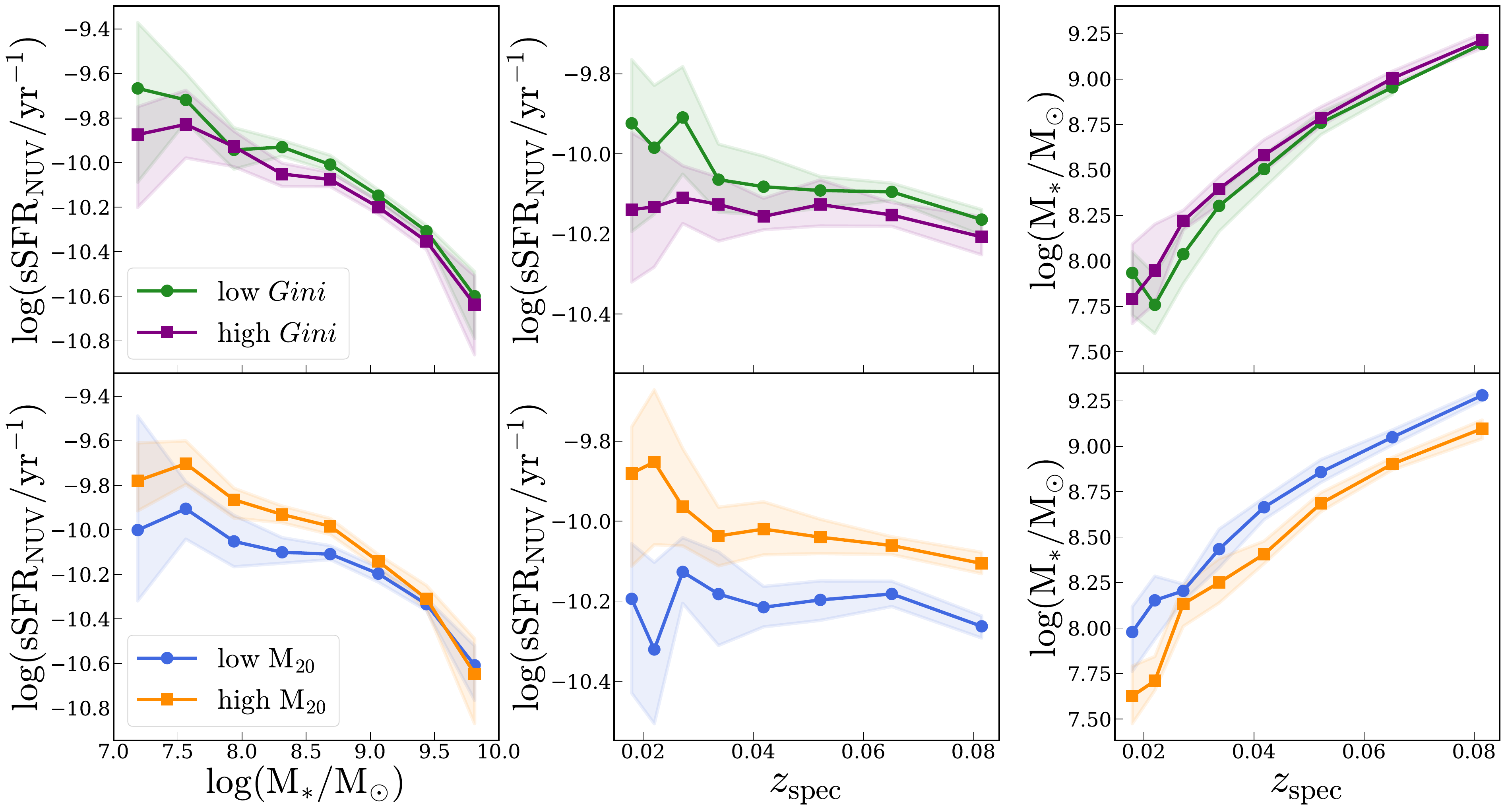}
\caption{Median trends in galaxy properties and redshift split by morphologies measured in $r$-band. The median values were evaluated through bootstrap resampling in each bin and the shaded regions represent the $3\sigma$ confidence interval of this distribution. Galaxies are divided into low and high values of $Gini$ (\textit{top row: green and purple}) and $M_{20}$ (\textit{bottom row: blue and orange}) based on median splits. \textit{Left column:} sSFR-$M_{\ast}$ plane, showing that galaxies with high $Gini$ and low $M_{20}$, i.e., more concentrated morphologies, have systematically lower sSFR at fixed stellar mass, consistent with Fig. \ref{fig:NPSplineSSFR}. \textit{Middle column:} sSFR-$z_{\rm spec}$ plane demonstrating mild evolution with cosmic time but persistent morphological differences at all redshifts. \textit{Right column:} $M_{\ast}-z_{\rm spec}$ plane in tune with Fig. \ref{fig:NPSplineM} that high $Gini$ and low $M_{20}$ occur at higher masses. \label{fig:zSFMS}}
\end{figure*}

The trends between the morphological measures and the galaxy properties that we have found in the preceding sections needs to be analyzed further because the stellar masses, star-formation rates and redshifts of the galaxies are mutually dependent. This will clarify if the morphologies, i.e., bulge prominence, of the SAGAbg low-mass galaxies truly scale with respect to $M_{\ast}$ and sSFR, evolve with $z_{\rm spec}$, or merely show trends that arise due to the selection effects. We proceed to do this by plotting the median trends of the SAGAbg-morph sample in the sSFR-$M_{\ast}$, sSFR-$z_{\rm spec}$ and $M_{\ast}-z_{\rm spec}$ planes in the three panels of Fig. \ref{fig:zSFMS}. We consider only the $r$-band measures in this analysis because the largest number of galaxies, i.e., 3947, pass the quality checks in Sec \ref{sec:diagnostics} in this band. The upper and lower rows correspond to the quantities $Gini$ and $M_{20}$ respectively. In order to illuminate the dependence of the galaxy morphologies we separate the measurements that are greater and lesser than the median measurements into `high' and `low' bins respectively. These are represented using the purple and green profiles for $Gini$ and orange and blue for $M_{20}$ respectively. We perform bootstrap resampling to sample the distributions of the `high' and `low' measures in bins of $M_{\ast}$ and $z_{\rm spec}$. From these distributions, we evaluate the medians along with the $3\sigma$ confidence intervals.

We find that the profiles of the larger and smaller values of the morphology measurements in each of the three planes diverge, with this being more pronounced for $M_{20}$ over $Gini$. A low $M_{20}$ or high Gini coefficient both imply stronger bulges. In the leftmost panel showing $M_{\ast}$-sSFR plane we find that the SAGAbg galaxies with lower values of $M_{20}$ have low sSFR overall, which is consistent with our inference from Fig. \ref{fig:NPSplineSSFR}. Both the profiles show decreasing trends with $M_{\ast}$ but we notice that at higher masses ${\rm log}(M_{\ast}/M_{\odot})\gtrsim 9$, the two profiles converge. This means that although we find a gradient in $M_{20}$ and $Gini$ that depends on sSFR at lower masses, at higher masses and likewise lower sSFR, the same divergence in the morphologies is absent.

The sSFR-$z_{\rm spec}$ plane in the center panel show the same divergence between high and low $M_{20}$ values and less significantly so for $Gini$. This is a result of their dependence on sSFR as we found in Fig. \ref{fig:NPSplineSSFR}. Both of the profiles show a gently decrease across $z_{\rm spec}$. However, the divergence in sSFR across the morphologies exists at all redshifts. This shows that the existence of the sSFR-morphology connection is independent of redshift at $z_{\rm spec}<0.1$. In the rightmost panel depicting the $M_{\ast}-z_{\rm spec}$ plane we find that lower values of $M_{20}$ have higher masses at all redshifts. This is again consistent with the median trend in Fig. \ref{fig:NPSplineM}. The divergence in the high and low $M_{20}$ profiles exist at all redshifts similar to the sSFR-$z_{\rm spec}$ plane, shows that dependence of $M_{20}$ on stellar mass is present across $z_{\rm spec}<0.1$.

\subsection{Dimensionality Reduction} \label{sec:umap}

\begin{figure*}
\centering
\includegraphics[scale=0.4]{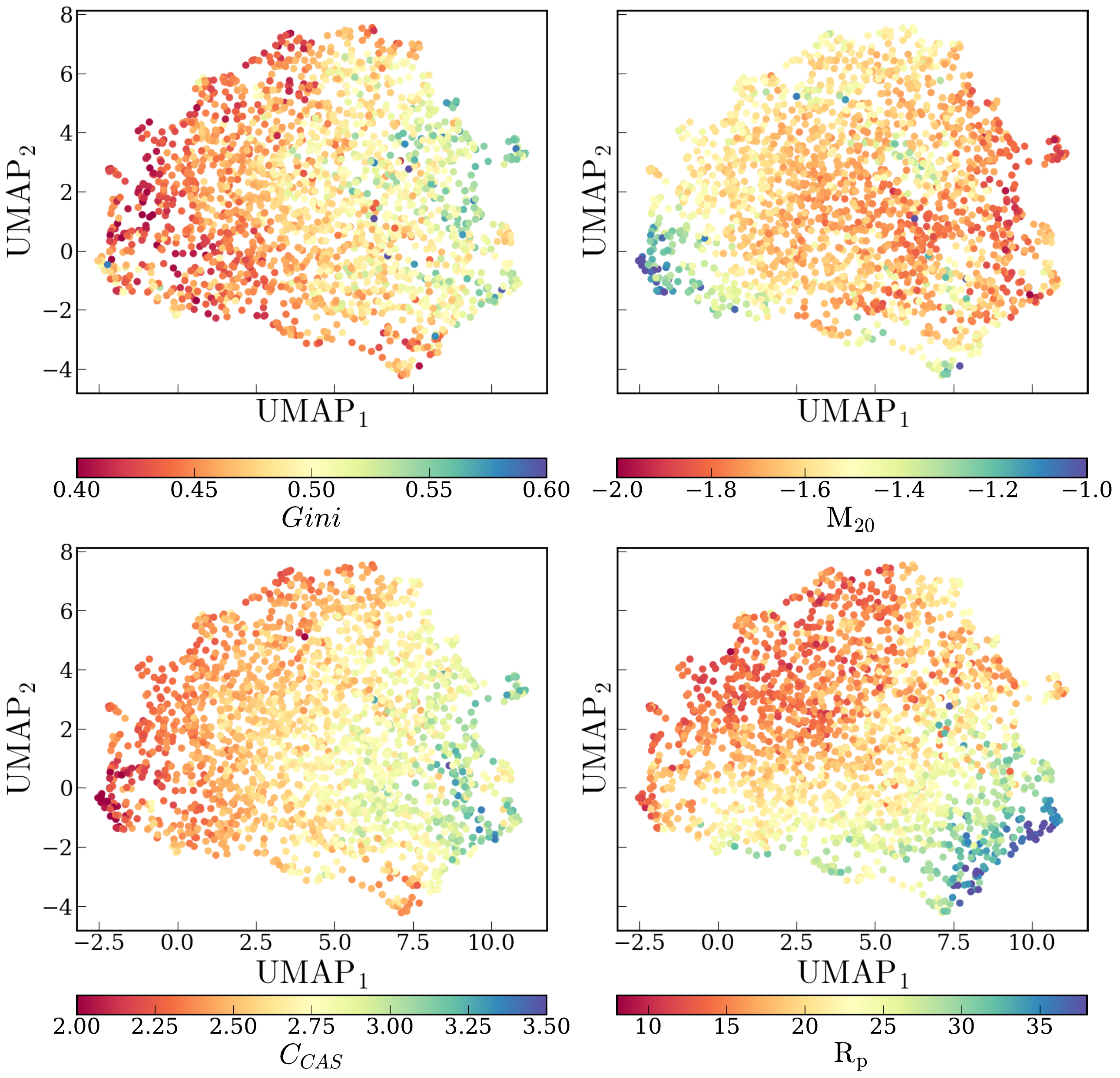}
\caption{UMAP embeddings based on the galaxy properties and the $gr$ band morphologies. Each panel shows the same two-dimensional UMAP embedding of galaxies, with points color-coded by different morphological parameters (\textit{clockwise from top-right}): $M_{20}$, Petrosian radius $R_p$, concentration $C_{CAS}$, and Gini coefficient. The smooth color gradients demonstrate that galaxies occupy a coherent manifold in the UMAP space. Compact, concentrated galaxies with high Gini coefficients (blue in upper-left panel) and low $M_{20}$ values (red in upper-right panel) occupy the same region, while extended, diffuse systems populate the lower-left of the distribution. The UMAP embedding effectively captures the correlations between different morphological indicators and the continuous nature of galaxy morphologies. \label{fig:umap}}
\end{figure*}

We have extracted a high-dimensional dataset of the galaxy morphologies from their multi-band imaging and this provides us the opportunity to probe patterns that are difficult to reveal in conventional analysis as undertaken above. We do this through the dimensionality reduction technique of Uniform Manifold Approximation and Projection for Dimension Reduction (UMAP)
\citep{2018arXiv180203426M}. We choose this method over alternatives Principal Component Analysis (PCA), t-distributed stochastic neighbor embedding (t-SNE) given its speed and efficiency. This nonlinear dimensionality reduction technique has been employed on complex morphology data of galaxies in literature \citep[e.g.][]{2023A&A...671A..19R,2025MNRAS.537..876A}.

We combine the morphology estimates from STATMORPH in the $gr$ bands, using those which satisfy the data-quality flags discussed in Sec. \ref{sec:diagnostics}. We augment this array with the $z_{\rm spec},{\rm log}(M_{\ast}/M_{\odot}),{\rm log}({\rm sSFR/yr}^{-1})$  data-vectors that link the morphology estimates with the respective galaxy properties. This produces a data-array of 2085 rows and 17 features. After scaling the data we apply the UMAP framework and assign appropriate values to the parameters. The \texttt{min\_dist} parameter corresponds to the minimum distance between points in the low-dimensional representational space thereby controlling the fine-level clustering while the \texttt{n\_neighbors} controls the number of nearest-neighbor points or alternatively the global structure of the embedding. We generated the low-dimensional embedding with different values of the parameters and finally chose the values ${\tt min\_dist}=0.4$ and ${\tt n\_neighbors}=5$ since this provided an appropriate balance between the fine and global levels of distribution in the low-dimensional UMAP space. 

We plot the UMAP embeddings in Fig. \ref{fig:umap} with the four-different sub-plots that are color-coded by the different morphological measures that are clockwise from top-right: $M_{20}$, Petrosian radius $R_p$, concentration $C_{CAS}$, and Gini coefficient. Once again we discount the $A_{CAS}$ and $S_{CAS}$ values on account of their large measurement uncertainties. We find the color gradients of $Gini$ and $C_{CAS}$ are very similar to each other and the gradient of $M_{20}$ in its opposite sense. This shows that the embedding space has resolved the bulge prominence of the SAGAbg-morph galaxies along this direction. The color gradient of $R_p$ howevers runs in a direction roughly orthogonal to the sense of $Gini$ and $C_{CAS}$, showing that the sizes of the galaxies run approximately independent of their bulge prominences.

\section{Discussion} \label{sec:discussion}

\subsection{Morphology \& Star Forming Main-Sequence}

Our work shows that the optical morphologies of the SAGAbg-morph galaxies at redshifts $z_{\rm spec}<0.1$ in terms of the $Gini-M_{20}$ distribution correspond to the late-type or the Sb/Sc/Irr class. Here we assume the classification scheme of \citet{2008ApJ...672..177L} defined for a high-mass sample of galaxies, holds for the less massive galaxies in our sample too. Nonetheless, we observe that these star forming dwarf galaxies have higher values of $M_{20}$, i.e., are disk-dominated, at low stellar mass and high sSFR and this effect is most prominent in $g$-band light. We place this in context with findings that claim dwarf galaxies have thick, puffy disks as opposed to the thin, regular disks in massive galaxies \citep{2008MNRAS.388.1321P,2013MNRAS.436L.104R}. According to \citet{2020ApJ...900..163K} dwarfs exhibit thick disk-like structures out up to $R_{\rm eff}$, with the blue galaxies being more disky than their red counterparts. The stellar feedback that causes the puffy disks in low-mass galaxies \citep{2010MNRAS.406L..65S}, becomes significant at $M_{\ast} \lesssim 10^9 M_{\odot}$ \citep{2020MNRAS.493.4126D,2025MNRAS.544.4651B}.

The bulge prominence traced by $M_{20}$ correlate with their primary properties- stellar mass and sSFR. We find that the disk galaxies with more concentrated light profiles have lower sSFR and higher $M_{\ast}$. The median trends in $M_{20}$ and $C_{CAS}$ as noted in our work are also consistently seen across the same mass range in both HSC observations and the TNG50 simulation in \citet{Martin2025cosmic}. In Fig. \ref{fig:sfms} the number of quenched systems increase notably at ${\rm log}(M_{\ast}/M_{\odot})\gtrsim 9$ and we can assume these galaxies also have strong bulges. While it has been demonstrated for more massive galaxies in this mass regime \citep[e.g.][]{2022MNRAS.513..256D,2025MNRAS.541.1164S,2025ApJ...985..206Z}, we show that this effect persists down to lower masses and is also connected to star formation activity.

In the regime of the SAGAbg-morph sample, this points to the formation of a pseudo-bulge. The fact that we find the bulge-strength strongly correlating with sSFR than stellar mass in Fig. \ref{fig:NPSplineSSFR} associates pseudo-bulges or the lack thereof to star formation activity   \citep{2006ApJ...642L..17F}. Pseudo-bulges can arise in star-forming disks due to secular processes \citep{2004ARA&A..42..603K,2007ApJ...664..640D,2009MNRAS.393.1531G} bar-driven gas inflows and internal disk instabilities. \citet{2025arXiv251025383B} show that stellar bulges can rapidly form at time-scales $< 1$ Gyr due to vigorous star-formation. This leads to bulges in galaxies of type Sb/Sc at $M_{\ast} \gtrsim 10^9 M_{\odot}$  \citep{2011ApJ...733L..47F} producing lower S{\'e}rsic indices $n < 2$ \citep{2008AJ....136..773F}. Although our trends suggest a gradual transition from disk to bulge-prominent morphologies, this could merely be an artifact of the uncertainties on mass ($\sim 0.2$ dex) and SFR ($\sim 0.1$ dex) dampening out a sharp transition at a certain mass or sSFR if it were to exist.



We also highlight in this work the explicit connection between the star-forming sequence (SFS) and morphologies of the dwarfs. We have shown how the gradients that exist across the distributions in the planes of sSFR-$M_{\ast}$ and $Gini-M_{20}$ are connected each other in Sec. \ref{sec:sfms_ev}. While \citet{Asali2025saga6} show by studying the SAGA satellites that the S{\'e}rsic index distributions of quenched and star-forming galaxies are similar, a increasing trend of the indices w.r.t stellar mass is found for both classes in Fig. 5. This once again is an example that bulge-strength increases with mass and that this effect is more pronounced for quenched dwarfs than star-forming dwarfs.



In summary, we have focused on the bulge prominence, i.e., concentration of the galaxy's light profile, due to the robustness of the metrics. For a further understanding of the morphology- star formation connection we need to also consider the sizes and environments of the galaxies. This is because the environment can still regulate the sizes of low-mass galaxies. \citet{Asali2025saga6} show how the SAGA satellite sample show a radius-mass relation that is different from their isolated counterparts. In contrast studying extremely isolated dwarfs in the void \citep{2023ApJ...951...52D} show that these galaxies can puff up, i.e., become dispersion supported as opposed to rotation supported, through internal processes without the need for environmental factors. In a future work we wish to incorporate the effects of environmental processes on the morphologies of the SAGAbg-morph galaxies as well.



\subsection{Systematics}

Our work is an attempt to study the morphologies of a large sample of low-redshift, low-mass galaxies in multiband $griz$ photometry. Doing so, we highlight both the merits and shortcomings of such an approach. While the ground-based Legacy Survey ensures wide coverage, atmospheric seeing and sky brightness effects contribute to degradation in image quality. We have mitigated this bias by employing a stringent quality check in Sec. \ref{sec:diagnostics}. Nonetheless, only the metrics that correspond to the concentration of the galaxy's light profile, i.e., $C_{CAS}, M_{20}$, show the least fractional uncertainty in each of the bands and greater values of $MI$ score to be deemed robust for our analysis (see Sec. \ref{sec:information}). $Gini$ indices have very low $MI$ scores comparable to $A_{CAS}$ and $S_{CAS}$. When analyzing the $Gini,M_{20}$ trends and the joint distributions in Sec. \ref{sec:trends} and \ref{sec:GiniM20_space} it becomes apparent that between the metrics, the $Gini$ of more massive, distant galaxies are prone to biases arising from the low S/N and low resolution \citep{Sazonova2025statmorph} of the images. Meanwhile, $M_{20}$ is weighted strongly by pixels close to the galaxy's center as opposed to $Gini$ and is only biased by low resolution. This is why we place the systematic trends of $M_{20}$ as a measure of bulge prominence, on a stronger footing as compared to $Gini$.

We are aware of the consequences of dust attenuation on the morphology measures in this work. This results in dimming and reddening of the spectra at bluer wavelengths \citep{2000ApJ...533..682C}. The impact of dust on the optical morphology traces the underlying metallicity gradient which is observed to be more significant closer to the galaxy's center \citep{2004A&A...424..465B}. We expect that dwarf galaxies contain low dust content and low metallicity compared to their massive counterparts \citep{2013ApJ...779..102K}. Moreover, the metallicity gradient for dwarfs has been measured to be not significant \citep{2009ApJ...705..723C}. JWST MIRI observations in low-mass galaxies in \citet{2026arXiv260209085K} show their dust content is mainly in the form of dust clumps and dust lanes. Therefore we can assume that the effect of dust on the bulge strength measures primarily studied in this work should be negligible. \citet{2024ApJ...974..273M} demonstrate that the bias is at most 10-20\% for the highest-mass bin ($ 10^{9.8} M_{\odot} < M_{\ast} < 10^{10.3} M_{\odot}$) for $Gini,M_{20}$ measurements in the H$\alpha$. In light of these findings, we assume that dust does not significantly bias our measurements.

However, a limitation of our work has been the inability to reliably measure the disturbance and smoothness statistics that include $A_{CAS}$ and $S_{CAS}$. \citet{Sazonova2024rms} show that the average sky background term $A_{\rm bgr}$ is biased due to depth and resolution. This leads to over-subtraction of the background asymmetry and subsequent under-estimation of $A_{CAS}$ values. $S_{CAS}$ depends strongly on S/N, and can fluctuate about zero when low depth imaging is used \citep{2019Rodriguez-Gomez}. As a result, we are unable to study inhomogeneities in the light distribution that would be caused due to features including star-forming regions, tidal streams and minor mergers in this work, even though these features are found in visual inspection of the cutouts. Therefore to probe the inhomogeneities in the light distribution of the SAGAbg galaxies we require deeper photometry that we propose as a future work.


\section{Conclusion}

We comprehensively study the morphologies of galaxies at redshifts $z_{\rm spec}<0.1$ in the SAGAbg catalog. We apply a magnitude selection $18<m_r<20.75$ to select low-mass galaxies $7\lesssim {\rm log}(M_{\ast}/M_{\odot})\lesssim 10$. The sample of 6211 galaxies are star-forming and predominantly located in the field. We complement our data with star-formation rates derived from GALEX NUV band measurements. We perform segmentation and deblending on $griz$ band cutouts from Legacy Survey cutouts of these sources and use the STATMORPH code to estimate non-parametric measures of morphology that include the Gini index, $M_{20}$, $CAS$ parameters and S{\'e}rsic model fits. We statistically analyze the morphological measures and find:

\begin{itemize}
    \item The measures of bulge strength- $M_{20},C_{CAS},Gini$, i.e., the concentration of the galaxy's light profiles, are the most reliable in terms of low fractional uncertainties and high mutual information scores. The low $\langle \text{S/N}\rangle$ imaging results in nonphysical values of $A_{CAS},S_{CAS}$, Petrosian radius and S{\'e}rsic model parameters.
    
    \item Median values of $M_{20}$ that trace bulge prominence, decrease with stellar mass and increase with sSFR across all bands with the trends of the $C_{CAS}$ parameter also in agreement. The $g$-band morphologies particularly indicate the weakest bulges implying flatter, disk morphologies in bluer light.
    
    \item In the Gini-$M_{20}$ space, the sample occupies the empirical region that corresponds to galaxies with Sb/Sc/Ir morphologies. A sequence is observed in this space with clustering that depends on galaxy properties, e.g., stellar mass, and can be linked to the star-forming sequence in the sSFR-$M_{\ast}$ plane.

    \item The centroid of the bivariate Gaussian function fit to the Gini-$M_{20}$ distribution shift with varying stellar mass and sSFR while any redshift evolution is found to be insignificant. After taking into account biases due to resolution and $\langle \text{S/N}\rangle$, only the trends of mean $M_{20}$ are deemed robust and are consistent with the median trends observed above.

    \item The variation of the bulge strength with stellar mass and sSFR is an evidence of a transition from disks to pseudo-bulges across the mass scale we have considered. This may arise due to internal processes that are found in the quenched galaxies at the higher mass end of our sample.

\end{itemize}

The Legacy Survey of Space and Time (LSST) \citep{2019ApJ...873..111I} is set to provide improved photometry in terms of both survey depth and PSF width. This will further revolutionize our understanding of the morphologies of low-mass galaxies in the low-redshift universe. These vast datasets will warrant the use of machine learning models for a comprehensive analysis. However non-parametric morphologies, are powerful as we show in this work, offering direct insights on the physical processes shaping the galaxies.

\section{Acknowledgment}

JB is grateful to Marla Geha for feedback that enhanced key figures. JB thanks Eddie Schlafly for a clarification regarding the DECaLS inverse-variance maps. JB and AHGP acknowledge support from NSF grant No. AST-2008110 and AHGP also acknowledges support from NSF grant No. AST-2510899. MAdlR and JB acknowledge support from the Research Corporation for Scientific Advancement Scialog grants No. SA-LSST-2025-108c and SA-LSST-2025-121a. 

This research used data from the SAGA Survey (Satellites Around Galactic Analogs; sagasurvey.org). The SAGA Survey is a galaxy redshift survey with spectroscopic data obtained by the SAGA Survey team with the Anglo-Australian Telescope, MMT Observatory, Palomar Observatory, W. M. Keck Observatory, and the South African Astronomical Observatory (SAAO). The SAGA Survey also made use of many public data sets, including: imaging data from the Sloan Digital Sky Survey (SDSS), the Dark Energy Survey (DES), the GALEX Survey, and the Dark Energy Spectroscopic Instrument (DESI) Legacy Imaging Surveys, which includes the Dark Energy Camera Legacy Survey (DECaLS), the Beijing-Arizona Sky Survey (BASS), and the Mayall z-band Legacy Survey (MzLS); redshift catalogs from SDSS, DESI, the Galaxy And Mass Assembly (GAMA) Survey, the Prism Multi-object Survey (PRIMUS), the VIMOS Public Extragalactic Redshift Survey (VIPERS), the WiggleZ Dark Energy Survey (WiggleZ), the 2dF Galaxy Redshift Survey (2dFGRS), the HectoMAP Redshift Survey, the HETDEX Source Catalog, the 6dF Galaxy Survey (6dFGS), the Hectospec Cluster Survey (HeCS), the Australian Dark Energy Survey (OzDES), the 2-degree Field Lensing Survey (2dFLenS), and the Las Campanas Redshift Survey (LCRS); HI data from the Arecibo Legacy Fast ALFA Survey (ALFALFA), the FAST all sky HI Survey (FASHI), and HI Parkes All-Sky Survey (HIPASS); and compiled data from the NASA-Sloan Atlas (NSA), the Siena Galaxy Atlas (SGA), the HyperLeda database, and the Extragalactic Distance Database (EDD). The SAGA Survey was supported in part by NSF collaborative grants AST-1517148 and AST-1517422 and Heising–Simons Foundation grant 2019-1402. SAGA Survey's full acknowledgments can be found at https://sagasurvey.org/ack/. This research uses services or data provided by the Astro Data Lab, which is part of the Community Science and Data Center (CSDC) Program of NSF NOIRLab. NOIRLab is operated by the Association of Universities for Research in Astronomy (AURA), Inc. under a cooperative agreement with the U.S. National Science Foundation.

The Legacy Surveys consist of three individual and complementary projects: the Dark Energy Camera Legacy Survey (DECaLS; Proposal ID 2014B-0404; PIs: David Schlegel and Arjun Dey), the Beijing-Arizona Sky Survey (BASS; NOAO Prop. ID 2015A-0801; PIs: Zhou Xu and Xiaohui Fan), and the Mayall z-band Legacy Survey (MzLS; Prop. ID 2016A-0453; PI: Arjun Dey). DECaLS, BASS and MzLS together include data obtained, respectively, at the Blanco telescope, Cerro Tololo Inter-American Observatory, NSF’s NOIRLab; the Bok telescope, Steward Observatory, University of Arizona; and the Mayall telescope, Kitt Peak National Observatory, NOIRLab. Pipeline processing and analyses of the data were supported by NOIRLab and the Lawrence Berkeley National Laboratory (LBNL). The Legacy Surveys project is honored to be permitted to conduct astronomical research on Iolkam Du’ag (Kitt Peak), a mountain with particular significance to the Tohono O’odham Nation. LBNL is managed by the Regents of the University of California under contract to the U.S. Department of Energy. This project used data obtained with the Dark Energy Camera (DECam), which was constructed by the Dark Energy Survey (DES) collaboration. Funding for the DES Projects has been provided by the U.S. Department of Energy, the U.S. National Science Foundation, the Ministry of Science and Education of Spain, the Science and Technology Facilities Council of the United Kingdom, the Higher Education Funding Council for England, the National Center for Supercomputing Applications at the University of Illinois at Urbana-Champaign, the Kavli Institute of Cosmological Physics at the University of Chicago, Center for Cosmology and Astro-Particle Physics at the Ohio State University, the Mitchell Institute for Fundamental Physics and Astronomy at Texas A\&M University, Financiadora de Estudos e Projetos, Fundacao Carlos Chagas Filho de Amparo a Pesquisa do Estado do Rio de Janeiro, Conselho Nacional de Desenvolvimento Cientifico e Tecnologico and the Ministerio da Ciencia, Tecnologia e Inovacao, the Deutsche Forschungsgemeinschaft and the Collaborating Institutions in the Dark Energy Survey. The Collaborating Institutions are Argonne National Laboratory, the University of California at Santa Cruz, the University of Cambridge, Centro de Investigaciones Energeticas, Medioambientales y Tecnologicas-Madrid, the University of Chicago, University College London, the DES-Brazil Consortium, the University of Edinburgh, the Eidgenossische Technische Hochschule (ETH) Zurich, Fermi National Accelerator Laboratory, the University of Illinois at Urbana-Champaign, the Institut de Ciencies de l’Espai (IEEC/CSIC), the Institut de Fisica d’Altes Energies, Lawrence Berkeley National Laboratory, the Ludwig Maximilians Universitat Munchen and the associated Excellence Cluster Universe, the University of Michigan, NSF’s NOIRLab, the University of Nottingham, the Ohio State University, the University of Pennsylvania, the University of Portsmouth, SLAC National Accelerator Laboratory, Stanford University, the University of Sussex, and Texas A\&M University. BASS is a key project of the Telescope Access Program (TAP), which has been funded by the National Astronomical Observatories of China, the Chinese Academy of Sciences (the Strategic Priority Research Program “The Emergence of Cosmological Structures” Grant XDB09000000), and the Special Fund for Astronomy from the Ministry of Finance. The BASS is also supported by the External Cooperation Program of Chinese Academy of Sciences (Grant 114A11KYSB20160057), and Chinese National Natural Science Foundation (Grant 12120101003, 11433005). The Legacy Survey team makes use of data products from the Near-Earth Object Wide-field Infrared Survey Explorer (NEOWISE), which is a project of the Jet Propulsion Laboratory/California Institute of Technology. NEOWISE is funded by the National Aeronautics and Space Administration. The Legacy Surveys imaging of the DESI footprint is supported by the Director, Office of Science, Office of High Energy Physics of the U.S. Department of Energy under Contract No. DE-AC02-05CH1123, by the National Energy Research Scientific Computing Center, a DOE Office of Science User Facility under the same contract; and by the U.S. National Science Foundation, Division of Astronomical Sciences under Contract No. AST-0950945 to NOAO.


\software{\texttt{NumPy} \citep{2020Natur.585..357H}, \texttt{Matplotlib} \citep{2007CSE.....9...90H}, \texttt{SciPy} \citep{2020NatMe..17..261V}, \texttt{Astropy} \citep{2013A&A...558A..33A,2018AJ....156..123A}, PHOTUTILS \citep{larry_bradley_2025_14889440}, STATMORPH \citep{2019Rodriguez-Gomez}, \texttt{lmfit} \citep{2014zndo.....11813N}}

\bibliography{sample7}{}
\bibliographystyle{aasjournal}

\appendix

\section{Bivariate Gaussian Fitting in Gini-$M_{20}$ space} \label{sec:2dgaussian}

\begin{figure*}
\centering
\includegraphics[scale=0.4]{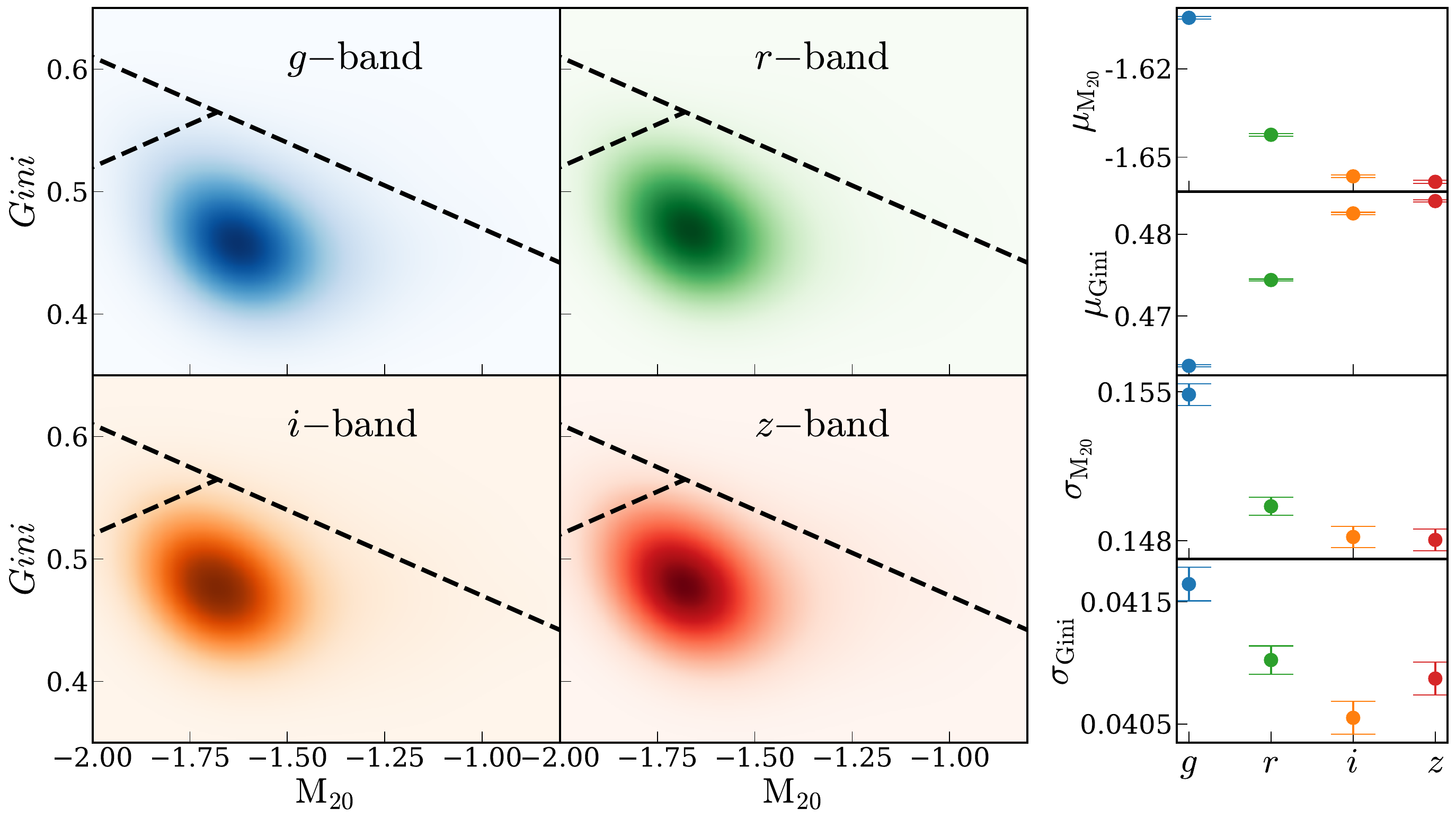}
\caption{Distribution of non-parametric morphological indicators- Gini coefficient and $M_{20}$ for galaxies observed in $griz$ bands. \textit{Left}: Two-dimensional Gaussian kernel density estimates (KDEs) showing the joint distribution of $Gini$ and $M_{20}$ values in each photometric band. The \textit{dashed-lines} represent Eq. \ref{eq:merger_line1}, \ref{eq:merger_line2}. \textit{Right}: Best-fit 2D Gaussian parameters derived from the observed distributions, displaying the mean values ($\mu_{\rm Gini}, \mu_{M_{20}}$) and standard deviations ($\sigma_{\rm Gini}, \sigma_{M_{20}}$) for each band. \label{fig:GiniM20} }
\end{figure*}

The four subplots on the left side of Fig. \ref{fig:GiniM20} represent the 2D gaussian Kernel Density Estimates (KDE) in the $Gini-M_{20}$ planes for each band. The bandwidth for the gaussian KDE was chosen as 0.08 as this was found optimum, i.e. did not lead to overfitting or excess smoothing of the distribution.  Eq. \ref{eq:merger_line1} and \ref{eq:merger_line2} are plotted as dotted lines, with our sample lying on the sequence as shown by the over-density in darker colors. We notice that the ${\it griz}$ band KDEs in the $Gini-M_{20}$ spaces possess smooth and well-behaved contours. We fit 2D Gaussian functions to each of the KDEs using the \texttt{lmfit} package \citep{2014zndo.....11813N}.

\begin{equation}
    p(\mathbf{x}) = \frac{\exp\left(-\frac{1}{2}(\mathbf{x} - \boldsymbol{\mu})^{\mathrm{T}} \boldsymbol{\Sigma}^{-1} (\mathbf{x} - \boldsymbol{\mu})\right)}{2\pi\sqrt{|\boldsymbol{\Sigma}|}}.
\end{equation}

In this case, the distribution is parameterized by the means- $\mu_{\rm Gini}, \mu_{M_{20}}$ and variances- $\sigma_{\rm Gini}^2, \sigma_{M_{20}}^2$, representative of the centers and widths of the sequence in $Gini-M_{20}$ space respectively. These in addition to the correlation coefficient $\rho$ together make up the mean and covariance arrays as,

\begin{equation}
   \boldsymbol{\mu} = \begin{pmatrix} \mu_{M_{20}} \\ \mu_{\rm Gini} \end{pmatrix}, \quad \boldsymbol{\Sigma} = \begin{pmatrix} \sigma_{M_{20}}^2 & \rho\sigma_{M_{20}}\sigma_{\rm Gini} \\ \rho\sigma_{M_{20}}\sigma_{\rm Gini} & \sigma_{\rm Gini}^2 \end{pmatrix}.
\end{equation}

The best-fit parameters in the $griz$ bands are plotted in the right panel of Fig. \ref{fig:GiniM20} along with the error bars representing the $1\sigma$ uncertainty of the parameters. The variations among the bands for the means are $\sim 2\%$ with $\mu_{\rm Gini}$ and $ \mu_{M_{20}}$ both decreasing with increasing wavelength. If we assume that systematics due to dust or noise are not biasing our results, this means that while the concentration of the central bulge is increasing with wavelength as shown by a decrease in $\mu_{M_{20}}$, the overall light profile becomes more uniform as shown by a decrease in $\mu_{\rm Gini}$ since the Gini index is not weighted by pixels close to the center of the galaxy like $M_{20}$ does. Furthermore we find that the dispersions --- $\sigma_{\rm Gini}^2, \sigma_{M_{20}}^2$ --- get smaller from $g$ to $z$-band showing that the sequence gets tighter with increasing wavelength.

\section{Corrections to Gini-$M_{20}$ Distribution} \label{sec:correctGM20}

\begin{figure*}
\centering
\includegraphics[scale=0.275]{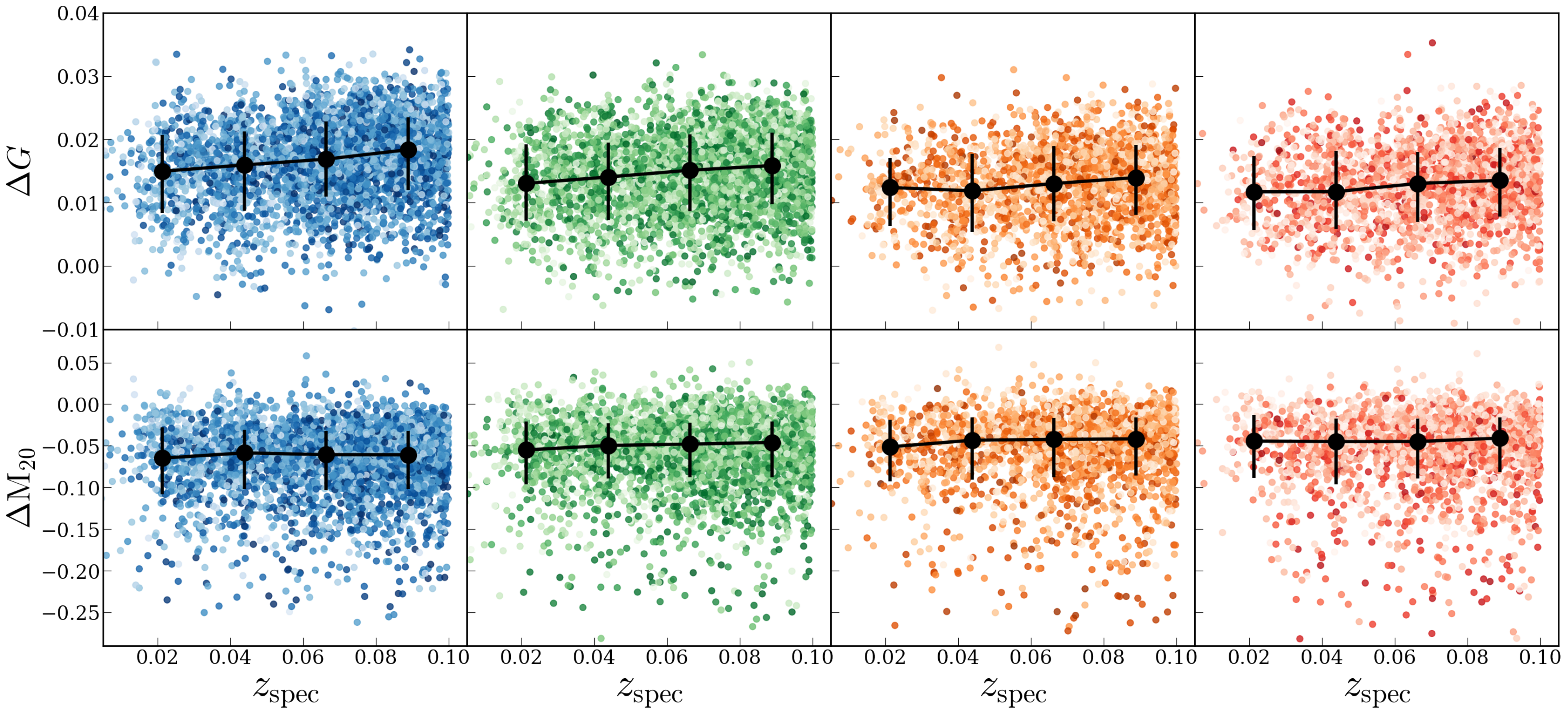}
\caption{Redshift evolution of the corrections on the Gini index and $M_{20}$ measures that were calculated using the presciption of \citet{Sazonova2025statmorph}. All the points here satisfy the quality flags outlined in Sec. \ref{sec:diagnostics} and are color-coded by apparent magnitude in each band. \textit{Top} row: Corrections on the Gini indices plotted against spectroscopic redshift ($z_{\rm spec}$). \textit{Bottom} row: Corrections on $M_{20}$ ($\mu_{M_{20}}$) versus spectroscopic redshift $z_{\rm spec}$. The \textit{black} error-bars show the median values along with the 16-84$^{\rm th}$ percentile in bins of $z_{\rm spec}$. This shows that $Gini$ is under-estimated by $\sim 0.015$ while $M_{20}$ is over-estimated by $\sim 0.05$ due to systematics in all bands at all redshifts. However the biases are usually constant with redshift, especially for $M_{20}$.
\label{fig:GM20_corrections}}
\end{figure*}

Here we test the robustness of the monotonic trend of the mean values $\mu_{M_{20}}$ and $\mu_{\rm Gini}$ with redshift that we noticed in Fig. \ref{fig:GM20_redshift}. To do this we apply the formalism outlined in \citet{Sazonova2025statmorph} where the effect of noise and resolution on the morphology estimates are studied. The work also provides prescriptions for correcting these estimates using the signal-to-noise per pixel $\langle\text{S/N}\rangle$ and the effective resolution $\mathcal{R}_{\text{eff}}$. While the former quantity is the same as discussed in Sec. \ref{sec:diagnostics}, $\mathcal{R}_{\text{eff}}$ is defined as the ratio of the Petrosian radius with respect to the pixel scale. Using symbolic regression \citet{Sazonova2025statmorph} determines the correction for the Gini index as,

\begin{equation}
G_{\text{corrected}} = G_{\text{measured}} \left(0.83 - \frac{0.59}{\mathcal{R}_{\text{eff}}}\right) + 0.1\tanh\langle\text{S/N}\rangle,
\end{equation}

and the correction for $M_{20}$ as,

\begin{equation}
M_{20,\text{corrected}} = M_{20,\text{measured}} \left(0.74 - \frac{1}{\mathcal{R}_{\text{eff}}}\right)  - 0.5.   
\end{equation}

In Fig. \ref{fig:GM20_corrections} we plot the corrections on the Gini indices and $M_{20}$ statistics that are $\Delta Gini = Gini_{\text{corrected}}-Gini_{\text{measured}}$ and $\Delta M_{20} = M_{20,\text{corrected}}-M_{20,\text{measured}}$ against the redshifts of the clean SAGAbg-morph sample in the $griz$ bands. The black error-bars represent the median values along with the 16-84$^{\rm th}$ percentile in bins of $z_{\rm spec}$. We find that while $Gini$ is under-estimated by $\sim 0.015$, $M_{20}$ is over-estimated by $\sim 0.05$, both corresponding to a correction of $\sim 3\%$. These biases are roughly constant across redshift, especially for $M_{20}$.

\end{document}